\documentclass[12pt, a4paper]{article}
\pdfoutput=1
\usepackage{jheppub}

\usepackage{amsmath}
\usepackage{amssymb}
\usepackage{multirow}
\usepackage{sidecap}
\usepackage{subfigure}
\usepackage{graphicx}
\usepackage{epstopdf}
\usepackage{dcolumn}
\usepackage{bm}

 \newcommand{\kin}[0]{{\mathcal{X}}}

\newcommand{\half}[0]{\frac{1}{2}}

\newcommand{\pd}[2]{\frac{\partial #1}{\partial #2}}
\newcommand{\ld}[0]{\mathcal{L}}

\newcommand{\dd}[0]{{\rm{d}}}
\newcommand{\defn}[0]{\equiv}

\newcommand{\qsubrm}[2]{{#1}_{\scriptscriptstyle{\textrm{#2}}}}
\newcommand{\qsuprm}[2]{{#1}^{\scriptscriptstyle\textrm{#2}}}
 \newcommand{\lcdm}[0]{$\Lambda$CDM }
\newcommand{\cs}[3]{\Gamma^{#1}_{\,\,\,\, #2#3}}

\newcommand{\ep}[0]{{ {\delta}_{\scriptscriptstyle{\rm{E}}}}}
\newcommand{\lp}[0]{{ {\delta}_{\scriptscriptstyle{\rm{L}}}}}

\newcommand{\rbm}[1]{{\bf{#1}}}

\newcolumntype{V}{>{\centering\arraybackslash} m{.4\linewidth} }
\newcommand{\hct}[0]{\mathcal{H}}
\newcommand{\subpsm}[3]{\subsm{#1}{#2}^{\scriptscriptstyle #3}}
\newcommand{\supsm}[2]{{#1}^{\scriptscriptstyle{#2}}}
 \newcommand{\subsm}[2]{{#1}_{\scriptscriptstyle{#2}}}

\newcommand{\AW}[0]{A_{\mathcal{W}}}
\newcommand{\BW}[0]{B_{\mathcal{W}}}
\newcommand{\CW}[0]{C_{\mathcal{W}}}
\newcommand{\DW}[0]{D_{\mathcal{W}}}
\newcommand{\EW}[0]{E_{\mathcal{W}}}

\newcommand{\lied}[1]{\pounds_{#1}}

\newcommand{\vphi}[0]{\delta\phi}
\newcommand{\sol}[0]{\Diamond^2\ld}

\newcommand{\nphiu}[1]{\nabla^{#1}\phi}
\newcommand{\nphid}[1]{\nabla_{#1}\phi}

\renewcommand{\figurename}{Figure}

\def\be{\begin{equation}}
\def\ee{\end{equation}}
\def\bea{\begin{eqnarray}}
\def\eea{\end{eqnarray}}
\def\bse{\begin{subequations}}
\def\ese{\end{subequations}}

\newcommand{\fref}[1]{{Figure \ref{#1}}}
\newcommand{\tref}[1]{{Table \ref{#1}}}

\let\oldsqrt\sqrt
\def\sqrt{\mathpalette\DHLhksqrt}
\def\DHLhksqrt#1#2{%
\setbox0=\hbox{$#1\oldsqrt{#2\,}$}\dimen0=\ht0
\advance\dimen0-0.2\ht0
\setbox2=\hbox{\vrule height\ht0 depth -\dimen0}%
{\box0\lower0.4pt\box2}}

\subheader{\today}

\title{Effective action approach to cosmological perturbations in dark energy and modified gravity}

\author[a]{Richard A. Battye}
\author[a]{and Jonathan A. Pearson}
\affiliation[a]{Jodrell Bank Centre for Astrophysics, School of Physics and Astronomy, The University of Manchester, Manchester M13 9PL, U.K}
\emailAdd{richard.battye@manchester.ac.uk}
\emailAdd{jp@jb.man.ac.uk}

\abstract{In light of upcoming observations modelling perturbations in dark energy and modified gravity models has become an important topic of research. We develop an effective action to construct the components of the perturbed dark energy momentum tensor which appears in the perturbed generalized gravitational field equations, $\delta G^{\mu\nu} = 8 \pi G \delta T^{\mu\nu} + \delta U^{\mu\nu}$ for linearized perturbations. Our method does not require knowledge of the Lagrangian density of the dark sector to be provided, only its field content. The method is based on the fact that it is only necessary to specify the perturbed Lagrangian to quadratic order and couples this with the assumption of global statistical isotropy of spatial sections to show that the model can be specified completely in terms of a finite number of background dependent functions.  We present our formalism in a coordinate independent fashion and provide explicit formulae for the perturbed conservation equation and the components of ${\delta U^{\mu}}_{\nu}$ for two explicit generic examples: (i) the dark sector does not contain extra fields, $\ld = \ld (g_{\mu\nu})$ and (ii) the dark sector contains a scalar field  and its first derivative $\ld = \ld(g_{\mu\nu},\phi, \nabla_{\mu}\phi)$.  We discuss how the formalism can be applied to modified gravity models   containing derivatives of the metric, curvature tensors, higher derivatives of the scalar fields and vector fields.}

\begin{document}
 \maketitle

\section{Introduction}
The standard model of cosmology uses General Relativity (GR) to describe gravitational interactions, an homogeneous/isotropic FRW metric to describe the geometry and matter content of cold dark matter (CDM)/photons/baryons to describe its constituents. Observations of the cosmic microwave background, supernovae, baryon acoustic oscillations, gravitational lensing and structure formation point to the existence of an additional component dubbed ``dark energy'',  or a modification to gravity, which needs to be introduced to explain the the observed acceleration \cite{1933AcHPh...6..110Z, Riess:1998cb, Perlmutter:1998np,  Riess:1998dv, 0067-0049-192-2-18}. 

The simplest explanation is a cosmological constant, $\Lambda$, and the standard paradigm is the \lcdm model. However, there is still considerable flexibility for the explanation to be something   radically different. In general,  we can model all possible theories as an extra ``dark sector" component to the stress-energy-momentum tensor. The structure of the gravitational field equations means that this extra component can be used to model either ``exotic matter''   with an equation of state $P/\rho<-{1\over 3}$ or a modification to GR (i.e. modifying exactly how gravity responds to the presence of matter). Constructing viable models of modified gravity has become an important task with the discovery of the acceleration of the Universe; some modified gravity models may also be able to account for observations which otherwise require dark matter.  

One way to model the dark sector is     ``Lagrangian engineering'': write down ever more complicated new theories  with a view of constraining their parameters and free functions to fit observation with the hope that self-accelerating solutions can be found. Theories where explicit forms of  dark energy are written down also fall into this category. They include  TeVeS \cite{PhysRevD.70.083509, Skordis:2009bf}, Einstein-\ae ther \cite{PhysRevD.75.044017}, Brans-Dicke \cite{PhysRev.124.925}, Horndeski \cite{springerlink:10.1007/BF01807638, Kobayashi:2011nu, Charmousis:2011bf} and $F(R)$ gravities \cite{Capozziello:2003tk, PhysRevD.70.043528}, quintessence \cite{PhysRevD.26.2580, Copeland:2006wr}, $k$-essence \cite{ArmendarizPicon:1999rj, ArmendarizPicon:2000ah} and Gallileons  \cite{PhysRevD.79.064036}. This is by no means an exhaustive list, and we have made no mention of the plethora of higher dimensional theories. The   reader is directed to the recent extensive review of modified gravity theories \cite{Clifton:2011jh}.

Given this proliferation of modified gravity and dark energy models, it would be a good idea to construct a generic way of parameterising deviations from the  GR+\lcdm picture and various suggestions have been made \cite{0004-637X-506-2-485, Weller:2003hw, PhysRevD.69.083503, PhysRevD.76.104043,   PhysRevD.77.103524, 1475-7516-2008-04-013,  PhysRevD.79.123527,  PhysRevD.81.083534, PhysRevD.81.104023,  Appleby:2010dx, Hojjati:2011ix, Baker:2011jy, Baker:2011wt} to do this for perturbations. This approach is called the ``Parameterized-Post-Friedmannian'' (PPF) framework, in analogy to the well established Parameterized-Post-Newtonian (PPN) framework which was invented for Solar System     tests of General Relativity \cite{will_PPN}. However, as we describe below, to date no generic approach has been proposed which has a physical basis.

In this paper we describe a new way of parameterizing perturbations in the dark sector requiring as an assumption knowledge of the field content. We do not assume a specific Lagrangian density, but we are able to model the possible effects on observations by using an effective action to compute the possible perturbations to the gravitational field equations. This is done by limiting the action to terms which are quadratic in the perturbed field content which is sufficient to model linearized perturbations,  and assuming that the spatial sections are isotropic.  In this paper we only consider the case that the dark sector contains first order derivatives of scalar fields and the metric; we will discuss higher order derivatives and vector fields in a follow-up paper.

Our theories will be completely general allowing for all possible degrees of freedom. Initially we do not impose reparametrization, or gauge, invariance. This is something which we would expect of a fundamental theory of dark energy, but not necessarily one for which the field content is just a coarse grained description. We will find that this can lead to an phenomological vector degree of freedom, $\xi^{\mu}$. In the elastic dark energy theory  \cite{Battye:2005mm, Battye:2006mb, PhysRevD.76.023005, Battye:2009ze}, which can be used to describe the effects of a dark energy component composed of a topological defect lattice, this represents a perturbation of the elastic medium from its equilibrium. We will see that the imposition of reparametrization invariance substantially reduces the number of free functions.

We note that many authors have consider possible dark energy theories which are effective Lagrangians in the traditional sense, that is, the terms in the Lagrangian represent an expansion of field operators which are suppressed at low energies  \cite{Weinberg:2008hq, Cheung:2007st, Creminelli:2008wc}. Our approach here is sufficiently similar to this approach to share the epitaph ``effective action", but it is completely different in many ways. It is completely classical and is in no sense an expansion energy scale. Moreover, it is just an effective action for the perturbations, and in no sense represents the full field theory of the dark energy.

\section{Approaches to parameterizing dark sector perturbations}

In this section we will provide a brief review of current approaches to studying generalized gravitational theories, concluding with a short discussion on the generalities of our approach. 

\subsection{Parameterized post-Friedmannian approach}
A popular way to parameterize the dark sector takes an ``observational'' perspective.  One can modify the equations governing the predictions of the Newtonian gravitational potential $\Phi$ and shear $\sigma$ by introducing extra functions space and time into the relevant equations and then parametrizing these extra functions in an {\it ad hoc} fashion. Since it is possible to explicitly observe $\Phi$ and $\sigma$  via the evolution structure and gravitational shear \cite{PhysRevD.69.083503, PhysRevD.76.023507, PhysRevD.81.083534, PhysRevD.81.104023, Hojjati:2011ix} (see also the more recent papers \cite{Dossett:2011tn, Kirk:2011sw, Laszlo:2011sv}), one can then compare them with the predictions of particular {\it ad hoc} choice and determine constraints on the deviation of a particular parameter from its value in General Relativity.

One way of doing this is  by modifying the  Poisson and gravitational slip equations, introducing two scale- and time-dependent functions, $Q = Q(k,a)$ and $R = R(k,a)$. The Poisson and gravitational-slip equations then become
\bea
k^2\Phi = - 4\pi G Qa^2\rho\Delta ,\qquad \Psi - R\Phi = - 12\pi G Qa^2 \rho(1+w)\sigma,
\eea
where $\Delta \defn \delta + 3 H\theta(1+w)$ is the comoving density perturbation, $\delta \defn \delta\rho/\rho$ the density contrast, $\theta$ the velocity divergence field, $w = P/\rho$ the equation of state and $\sigma$ is the anisotropic stress. When these equations are derived in GR one finds that $Q(k,a) = R(k,a) = 1$, and so if, by comparison to data, either of these parameters are shown to be inconsistent with unity, then deviations from GR can be established. In \cite{Baker:2011jy, Zuntz:2011aq} it was shown that the two functions $Q, R$ are not necessarily independent: they can be  linked by the perturbed Bianchi identity, depending on the structure of the underlying theory.

\subsection{Generalized gravitational field equations}
\label{skordis}
Another way to investigate the dark sector   takes a more theoretical standpoint, and is based on a more consistent modification of the governing field equations. The method stems from the fact that any modified gravity theory or model of dark energy can be encapsulated by writing the \textit{generalized gravitational field equations}
\bea
\label{eq:sec-1.1-gravfldeqns}
G_{\mu\nu} = 8\pi G T_{\mu\nu} + U_{\mu\nu},
\eea
where $G_{\mu\nu} $ is the Einstein tensor calculated from the spacetime metric, $T_{\mu\nu}$ is the energy-momentum tensor of all {\it known} species (radiation, Baryons, CDM etc) and $U_{\mu\nu}$ is a tensor which contains all {\it unknown} contributions to the gravitational field equations, which we call the \textit{dark energy-momentum tensor} \cite{PhysRevD.76.104043, PhysRevD.77.103524, PhysRevD.79.123527}.  

Because the Bianchi identity automatically holds for the Einstein tensor, $\nabla_{\mu}G^{\mu\nu} = 0$, in the standard case where the {\it known} and {\it unknown} sectors are   decoupled  (that is $\nabla_{\mu}T^{\mu\nu}=0$) we have the conservation law 
\bea
\nabla_{\mu} U^{\mu\nu}=0.
\eea
This represents a constraint equation on the extra parameters and functions that may appear in a parameterization of the dark sector at the level of the background. At perturbed order,  the parameterization of $\delta U^{\mu\nu}$ is constrained by the perturbed conservation law
\bea
\delta(\nabla_{\mu}U^{\mu\nu})=0.
\eea 
The shortcoming of  this approach is that one must supply   the components of $\delta U^{\mu\nu}$. Skordis \cite{PhysRevD.79.123527} does this by expanding the components $\delta {U^{\mu}}_{\nu}$ in terms of pseudo derivative operators acting upon gauge invariant combinations of metric perturbations, by imposing the principles that (a) the field equations remain at most second order and (b) the equations are gauge-form invariant. A particular form of these components were considered in \cite{PhysRevD.79.123527}:
\bse
\bea
- a^2\delta {U^0}_0 = \frac{1}{a}\mathcal{A} \hat{\Phi},
\quad
-a^2 \delta {U^0}_i = \nabla_i (\tfrac{1}{a^2}\mathcal{B}\hat{\Phi}),\quad
a^2\delta {U^i}_i = \mathcal{C}_1\hat{\Phi} + \mathcal{C}_2\dot{\hat{\Phi}} + \mathcal{C}_3 \hat{\Psi},
\eea
\bea
a^2\big[ \delta {U^i}_j - \tfrac{1}{3}\delta^i_j \delta {U^k}_k\big] = (\nabla^i\nabla_j - \tfrac{1}{3}\delta_{ij} \nabla^2)( \mathcal{D}_1\hat{\Phi} + \mathcal{D}_2\dot{\hat{\Phi}} + \mathcal{D}_3 \hat{\Psi}),
\eea
\ese
where $\mathcal{O} = \{\mathcal{A}, \mathcal{B}, \mathcal{C}_i, \mathcal{D}_i\}$ is a set of pseudo differential operators and $\{\hat{\Phi}, \hat{\Psi}\}$ are gauge invariant combinations of perturbed metric variables. The possible form that the elements of $\mathcal{O}$ can take is constrained by the perturbed Bianchi identity. For instance, it was shown that $\mathcal{C}_3 = \mathcal{D}_3 = 0$ is one of the sufficient consistency relations. A generalized version of this method can be found in \cite{Baker:2011jy, Clifton:2011jh}.

This scheme provides a way to compute and constrain observables without ever having to write down an explicit theory for the dark sector.  There  appears to be, however, a weakness in the current formulation of this strategy: there does not seem to be a   physically obvious way to interpret the $\mathcal{O}$; for example, if one were to find that  $\mathcal{C}_3=0$ is ``required'' for consistency with observational data, what does that impose physically upon the system? It is exactly this issue we address   in this paper.  

\subsection{Effective action approach}
The generalized gravitational field equations (\ref{eq:sec-1.1-gravfldeqns}) can be constructed from an action
\bea
 {S} = \int \dd^4x\, \sqrt{-g} \bigg[ R + 16\pi G \qsubrm{\ld}{m}-2\qsubrm{\ld}{d}\bigg].
\eea
The matter Lagrangian density $\qsubrm{\ld}{m}$ contains all known matter fields (e.g. baryons, photons) and is used to construct the known energy momentum tensor $T^{\mu\nu}$,  and the dark sector Lagrangian density $\qsubrm{\ld}{d}$ contains all ``unknown'' contributions to the gravitational sector,  and will be used to construct the dark energy momentum tensor $U^{\mu\nu}$. One can define
\bea
\label{eq:sec:defn-darkemt}
T^{\mu\nu}\defn \frac{2}{\sqrt{-g}}\frac{\delta}{\delta g_{\mu\nu}}( \sqrt{-g}\qsubrm{\ld}{m}),\qquad U^{\mu\nu} \defn - \frac{2}{\sqrt{-g}}\frac{\delta}{\delta g_{\mu\nu}} (\sqrt{-g}\qsubrm{\ld}{d}).
\eea
The dark sector Lagrangian may contain known fields in an unknown configuration or extra fields, but of course we do not know \textit{a priori} what the  dark sector Lagrangian density is.

Two simple cases are (i) a slowly-rolling minimally coupled scalar field   parameterized by a potential, $V(\phi)$, and (ii) a modified gravity model parameterized by a free function of the Ricci scalar, $F(R)$. There are restrictions on the form of both of these functions to achieve acceleration, but once they have been applied there is still considerable freedom in the choices of $V(\phi)$ and $F(R)$ and wide ranges of behaviour of the expansion history, $a(t)$, can be arranged for particular choices of the functions. One would expect this to be the case in any self consistent dark energy model compatible with FRW metric and therefore it might seen reasonable to make the assumption that the dark stress-energy-momentum tensor $U_{\mu\nu} = \rho u_{\mu}u_{\nu} + P\gamma_{\mu\nu}$  where $w(a)=P/\rho$ is in 1-1 correspondence with $a(t)$. The important question, which we are concerned with, is how to parametrize the perturbations   $\delta {U^{\mu}}_{\nu}$ in a general way based on some general physical principle. In this way our approach is similar to that discussed in section~\ref{skordis}

The overall ethos which we advocate is to write down an effective action, inspired by the approach that is taken in particle physics (see, e.g. \cite{PhysRevD.8.1226}) where, for example, the most general modifications to the standard model are written down for a given field content that are compatible with some assumed symmetry/symmetries. Then all the free coefficients are constrained by experiment. In our case, we will specify the field content of the dark sector, for example, scalar or vector fields, and write down a general quadratic Lagrangian density for the perturbed field variables which is sufficient to generate equations of motion for linearized perturbations. We will also make the assumption that the spatial sections are isotropic which   substantially reduces the number of free coefficients.

\section{Formalism}
\subsection{Second order Lagrangian}
The underlying principle behind our method is to write down a effective Lagrangian density for perturbed field variables. If our theory is constructed from a set of field variables $\{X^{\scriptscriptstyle(\rm A)}\}$, then we write each field variable as a linearized perturbation about some background value,
\bea
X^{\scriptscriptstyle(\rm A)} = \bar{X}^{\scriptscriptstyle(\rm A)} + \delta X^{\scriptscriptstyle(\rm A)}.
\eea
The action for the perturbed field variables $\{\delta X^{\scriptscriptstyle(\rm A)}\}$ is computed by integrating a Lagrangian density which is quadratic in the perturbed field variables. If there are ``N'' perturbed field variables,  the effective Lagrangian density for the perturbed field variables is given by
\bea
 {\ld}_{\rm{\scriptscriptstyle eff}}(\delta X^{\scriptscriptstyle(\rm C)} ) = \sum_{{\rm A}=1}^{\rm{N}}\sum_{{\rm B}=1}^{\rm{N}}  \mathsf{G}_{{\scriptscriptstyle\rm A B}}  \delta X^{\scriptscriptstyle(\rm A)} \delta X^{\scriptscriptstyle(\rm B)},
\eea
where $\mathsf{G}_{{\scriptscriptstyle\rm A B}}=\mathsf{G}_{{\scriptscriptstyle\rm A B}}(\bar{X}^{\scriptscriptstyle(\rm C)}) $ is a set of  arbitrary functions only depending   on   the background field variables; clearly, $\mathsf{G}_{{\scriptscriptstyle\rm A B}} = \mathsf{G}_{{\scriptscriptstyle\rm  BA}}$. To obtain the equation of motion of the perturbed field variables $\{\delta X^{\scriptscriptstyle(\rm A)}\}$ we must induce some variation in the $\{\delta X^{\scriptscriptstyle(\rm A)}\}$ and subsequently demand that $ {\ld}_{\rm{\scriptscriptstyle eff}}$ is independent  of these variations. If we vary the perturbed field variables with a variational operator $\hat{\delta} $,
\bea
 \delta X^{\scriptscriptstyle(\rm A)} \rightarrow \delta X^{\scriptscriptstyle(\rm A)} + \hat{\delta} (\delta X^{\scriptscriptstyle(\rm A)} ),
\eea
then the effective Lagrangian will vary according to $ {\ld}_{\rm{\scriptscriptstyle eff}} \rightarrow  {\ld}_{\rm{\scriptscriptstyle eff}} + \hat{\delta} {\ld}_{\rm{\scriptscriptstyle eff}}$,
where
\bea
\hat{\delta} {\ld}_{\rm{\scriptscriptstyle eff}}  = 2\sum_{{\rm A}=1}^{\rm{N}}\sum_{{\rm B}=1}^{\rm{N}}  \mathsf{G}_{{\scriptscriptstyle\rm A B}}  \delta X^{\scriptscriptstyle(\rm A)}\hat{\delta} (\delta X^{\scriptscriptstyle(\rm B)}).
\eea
The demand that the effective Lagrangian is independent of these variations is the statement that
\bea
\frac{\hat{\delta}}{\hat{\delta}(\delta X^{\scriptscriptstyle(\rm B)})}{\ld}_{\rm{\scriptscriptstyle eff}} =0,
\eea
that is,
\bea
\sum_{{\rm A}=1}^{\rm{N}}\sum_{{\rm B}=1}^{\rm{N}}  \mathsf{G}_{{\scriptscriptstyle\rm A B}}  \delta X^{\scriptscriptstyle(\rm A)}=0.
\eea
These equations provide the equations of motion of the perturbed field variables. We will now show how to obtain the effective action for perturbations by directly perturbing the background action.

We will   consider an action of the form
\bea
\label{eq:sec1.8-action-proto}
S = \int \dd^4x\, \sqrt{-g} \ld,
\eea
where $g $ is the determinant of the spacetime metric, $g_{\mu\nu}$, and $\ld$ is the Lagrangian density, which contains all fields in the theory. It will be useful to write the first and second variations of the action as
\bea
\label{eqLsecLinto-dsdds}
\delta S = \int \dd^4x\, \sqrt{-g}\Diamond \ld,\qquad \delta^2S  = \int \dd^4x\, \sqrt{-g} \Diamond^2\ld,
\eea
where ``$\Diamond$'' is a useful measure-weighted pseudo-operator   introduced in \cite{0264-9381-11-11-010, Battye:1998zk} and is defined by
\bea
\Diamond^n \ld \defn \frac{1}{\sqrt{-g}}\delta^n(\sqrt{-g}\ld).
\eea
We will only consider first perturbations of the field content of a theory. 
For the action (\ref{eq:sec1.8-action-proto}) we can use the well known result
\bea
\label{eq:sec:3.4-vary-metdet}
\frac{1}{\sqrt{-g}}\delta \sqrt{-g} = - \half g_{\mu\nu} \delta g^{\mu\nu} = + \half g^{\mu\nu}\delta g_{\mu\nu},
\eea
 to show that to quadratic order in the perturbations that the integrands in (\ref{eqLsecLinto-dsdds})  are given by
\bse
\bea
\Diamond \ld =    \delta\ld + \half \ld g^{\mu\nu}\delta g_{\mu\nu},
\eea
\bea
\label{eq:2.14b-d2ld}
\Diamond^2 \ld  &=& \delta^2\ld  +  g^{\mu\nu}\delta g_{\mu\nu} \delta\ld + \frac{1}{4}\ld \bigg( g^{\mu\nu} g^{\alpha\beta} - 2g^{\mu(\alpha}g^{\beta)\nu}  \bigg)\delta g_{\mu\nu} \delta g_{\alpha\beta}. 
\eea
\ese
We treat the integrand of the second variation of the action, i.e. $\Diamond^2\ld$, as the effective Lagrangian, $\qsubrm{\ld}{eff}$,  for linearized perturbations, and it is called the \textit{second order Lagrangian}. The final term of (\ref{eq:2.14b-d2ld}) is an effective mass-term for the gravitational fluctuations $\delta g_{\mu\nu}$ which is always present even when the field which constitutes the dark sector does not vary, i.e. when $\delta\ld = \delta^2\ld = 0$.   

Although we will be providing various explicit examples later on in the paper, we will briefly discuss how to write down $\Diamond^2\ld$ once the field content has been specified. If the field content is $\{X,Y\}$, then we write $\ld = \ld(X,Y)$, and then $\Diamond^2\ld$ is written down by writing all quadratic interactions of the perturbed fields with appropriate coefficients,
\bea
\Diamond^2\ld = A(t)\delta X\delta X + B(t)\delta X\delta Y + C(t) \delta Y \delta Y.
\eea
Notice that we have moved from having  complete ignorance of how the fields $X,Y$ combine to construct the Lagrangian density $\ld$ to only requiring 3 ``background'' functions, $A(t),B(t),C(t)$ to be able to write $\Diamond^2\ld$ down. Typically, we would expect these functions to be specified in terms of the scale factor $a(t)$.

The theories we consider   contribute to the gravitational field equations via the dark energy-momentum tensor, $U^{\mu\nu}$, which we define in the usual way, (\ref{eq:sec:defn-darkemt}).
The indices on the dark energy momentum tensor are symmetric  by construction,
\bea
U_{\mu\nu} = U_{\nu\mu} = U_{(\mu\nu)},
\eea
where tensor indices are symmetrised as $A_{(\mu\nu)}  = \half \big(A_{\mu\nu} + A_{\nu\mu}\big)$. The dark energy-momentum tensor above can be directly perturbed to give
\bea
\delta U^{\mu\nu} = - \half \bigg[\sum_{\scriptscriptstyle\rm{A}}\bigg(\delta X^{\scriptscriptstyle(\rm A)} \frac{1}{\sqrt{-g}}\frac{\delta}{\delta X^{\scriptscriptstyle(\rm A)}} \frac{\delta}{\delta g_{\mu\nu}}(\sqrt{-g}\ld) \bigg) + U^{\mu\nu}g^{\alpha\beta}\delta g_{\alpha\beta}\bigg],
\eea
where $\{\delta X^{\scriptscriptstyle(\rm A)}\}$ are the perturbed field variables. This can be written in a   more succinct way by using the second order Lagrangian,
\bea
\label{eq:Sec:per-u-prototype-formalismsection}
\delta U^{\mu\nu} = - \half \bigg[4\pd{( \Diamond^2\ld)}{(\delta g_{\mu\nu})}+ U^{\mu\nu} g^{\alpha\beta} \delta g_{\alpha\beta} \bigg].
\eea
Therefore, to obtain the gravitational contribution at perturbed order, due to our effective Lagrangian for perturbed field variables, one must   compute the derivative of the second order Lagrangian with respect to the perturbed metric.

The equations of motion for a field $X$ and its perturbation $\delta X$ are found by regarding $ \ld$ and $\Diamond^2\ld$ as the relevant Lagrangian densities. Explicitly, the equations of motion for the   field $X$ and its perturbation, $\delta X$, are respectively given by
\bea
\label{eq:sec-2.10-el-pert}
\partial_{\mu}\bigg(\frac{\partial\ld}{\partial( \partial_{\mu}X)}\bigg) - \frac{\partial\ld}{\partial X}=0,\qquad \partial_{\mu}\bigg(\frac{\partial(\Diamond^2\ld)}{\partial( \partial_{\mu}\delta X)}\bigg) - \frac{\partial(\Diamond^2\ld)}{\partial \delta X}=0.
\eea
The   equations of motion governing the perturbation to the metric, $\delta g_{\mu\nu}$, are given by the perturbed   gravitational field equations, 
\bea
\delta G_{\mu\nu} = 8\pi G \delta T_{\mu\nu} + \delta U_{\mu\nu}.
\eea
The perturbed conservation law for the dark energy-momentum tensor is 
\bea
\delta (\nabla_{\mu}U^{\mu\nu})=0,
\eea
which can be written as
\bea
\label{eq:sec:per-bi-proto}
\nabla_{\mu}\delta U^{\mu\nu} + \half \bigg[ U^{\mu\nu}g^{\alpha\beta} - U^{\alpha\beta}g^{\mu\nu} + 2 g^{\nu\beta}U^{\alpha\mu}\bigg]\nabla_{\mu}\delta g_{\alpha\beta}=0.
\eea

\subsection{Isotropic (3+1) decomposition}
We will impose isotropy of spatial sections on the background spacetime. The motivation for doing this is that our goal is to study perturbations about an FRW background. After imposing isotropy we are able to use an isotropic (3+1) decomposition to significantly simplify expressions. It is also possible to include anisotropic backgrounds as described in \cite{Battye:2006mb}.

We will foliate the 4D spacetime by 3D surfaces orthogonal to a time-like vector $u_{\mu}$, which is normalized via 
\bea
 u^{\mu}u_{\mu} = -1.
 \eea
This induces an embedding of a 3D surface in a 4D space. The 4D metric is $g_{\mu\nu}$ and the 3D metric is $\gamma_{\mu\nu}$, and they are related by
\bea
\gamma_{\mu\nu} = g_{\mu\nu} + u_{\mu}u_{\nu} .
\eea
The foliation implies that the time-like vector is orthogonal to the 3D metric,
\bea 
u^{\mu}\gamma_{\mu\nu} = 0.
\eea
The foliation induces a symmetric extrinsic curvature, $K_{\mu\nu} \defn \nabla_{\mu}u_{\nu}$, which is entirely spatial, $u^{\mu}K_{\mu\nu} = 0$. We can use this to deduce that $\nabla_{\mu}\gamma_{\alpha\beta} = 2 K_{\mu(\alpha}u_{\beta)}$.

A common application of the (3+1)  decomposition is to write down the only energy-momentum tensor compatible with the globally isotropic FRW metric,
\bea
T_{\mu\nu} = \rho u_{\mu}u_{\nu} + P\gamma_{\mu\nu}.
\eea 
There are only two ``coefficients'' used in the decomposition of the energy-momentum tensor: the energy-density $\rho$ and pressure $P$,
\bea
\rho = u^{\mu}u^{\nu}T_{\mu\nu} ,\qquad P = \frac{1}{3}\gamma^{\mu\nu}T_{\mu\nu}.
\eea

Writing a tensor as a sum over combinations of $u^{\mu}$ and  $\gamma_{\mu\nu}$ defines the isotropic (3+1) decomposition. We will now show how to   decompose tensors of higher rank.  For example, an isotropic vector is completely decomposed as
\bea
A^{\mu} = A u^{\mu},
\eea
where $A = A(t)$. Notice that before we imposed isotropy upon $A^{\mu}$ we would need 4 functions to specify all ``free'' components of $A^{\mu}$; by imposing isotropy we have reduced the number of ``free'' functions from $4 \rightarrow 1$. A symmetric rank-2 isotropic tensor is completely decomposed as
\bea
\label{eq:sec:3.21-thingy}
B_{\mu\nu} =   B_1u_{\mu}u_{\nu}+B_2 \gamma_{\mu\nu} = B_{\nu\mu},
\eea
where $B_1 = B_1(t), B_2 = B_2(t)$.
The time-like part of $B_{\mu\nu}$ is $B_1$ and the space-like part  is $B_2$. A   rank-3 tensor symmetric in its second two indices is completely decomposed as
\bea
C_{\lambda\mu\nu} = C_1u_{\lambda}\gamma_{\mu\nu} + C_2u_{\lambda}u_{\mu}u_{\nu} + C_3 \gamma_{\lambda(\mu}u_{\nu)}=C_{\lambda\nu\mu}.
\eea
This formalism can also be used to construct tensors which are entirely spatial. For example, a rank-4 tensor defined as
\bea
D_{\mu\nu\alpha\beta} = D_1\gamma_{\mu\nu}\gamma_{\alpha\beta} + D_2 \gamma_{\mu(\alpha}\gamma_{\beta)\nu},
\eea
is entirely spatial, a fact which is manifested by $u^{\mu}D_{\mu\nu\alpha\beta} = 0$, after one notes the symmetries in the indices $D_{\mu\nu\alpha\beta} = D_{(\mu\nu)(\alpha\beta)} = D_{\alpha\beta\mu\nu}$.   

The coefficients which appear in an isotropic decomposition can only have time-like derivatives. For the coefficients $B_1, B_2$ in (\ref{eq:sec:3.21-thingy}) we have
\bea
\nabla_{\mu}B_1 = -\dot{B}_1u_{\mu},\qquad \nabla_{\mu}B_2 =- \dot{B}_2u_{\mu},
\eea
where an overdot is used to denote differentiation in the direction of the time-like vector: $\dot{X} \defn u^{\mu}\nabla_{\mu}X$.

\subsection{Perturbation theory}
We will be making substantial use of perturbation theory in this paper, and so here we will take the time to   concrete the notation and terminology we use. A large portion of the technology we are about to discuss was developed, amongst other things, to model   relativistic elastic materials \cite{Carter21111972,  BF01645505, PhysRevD.7.1590,  1978ApJ2221119F, carterqunt_1977, Carter26081980,   ll_elast, PhysRevD.60.043505, Battye:2006mb, PhysRevD.76.023005,   Carter:1982xm,     Azeyanagi:2009zd, Fukuma:2011pr}; we will recapitulate the ideas and bring the technology into the language of perturbation theory to be used with a gravitational theory.

A quantity $Q$ is perturbed about a background value, $\bar{Q}$,  as $Q = \bar{Q} + \delta Q$. For example, the metric perturbed  about a background $\bar{g}_{\mu\nu}$ is written as
\bea
g_{\mu\nu}  = \bar{g}_{\mu\nu}+ \delta g_{\mu\nu}.
\eea
It is important to realize that the operation of index raising and lowering does not commute with the variation. For example, $\delta g_{\mu\nu} = - g_{\mu(\alpha}g_{\beta)\nu} \delta g^{\alpha\beta}$
for the metric and $\delta(\nabla^{\mu}\phi) = g^{\mu\nu} \nabla_{\nu}\delta\phi + \delta g^{\mu\nu} \nabla_{\nu}\phi$ for the derivative of a scalar field $\phi$.  

\begin{figure*}[!t]
      \begin{center}
{\includegraphics[scale=0.6]{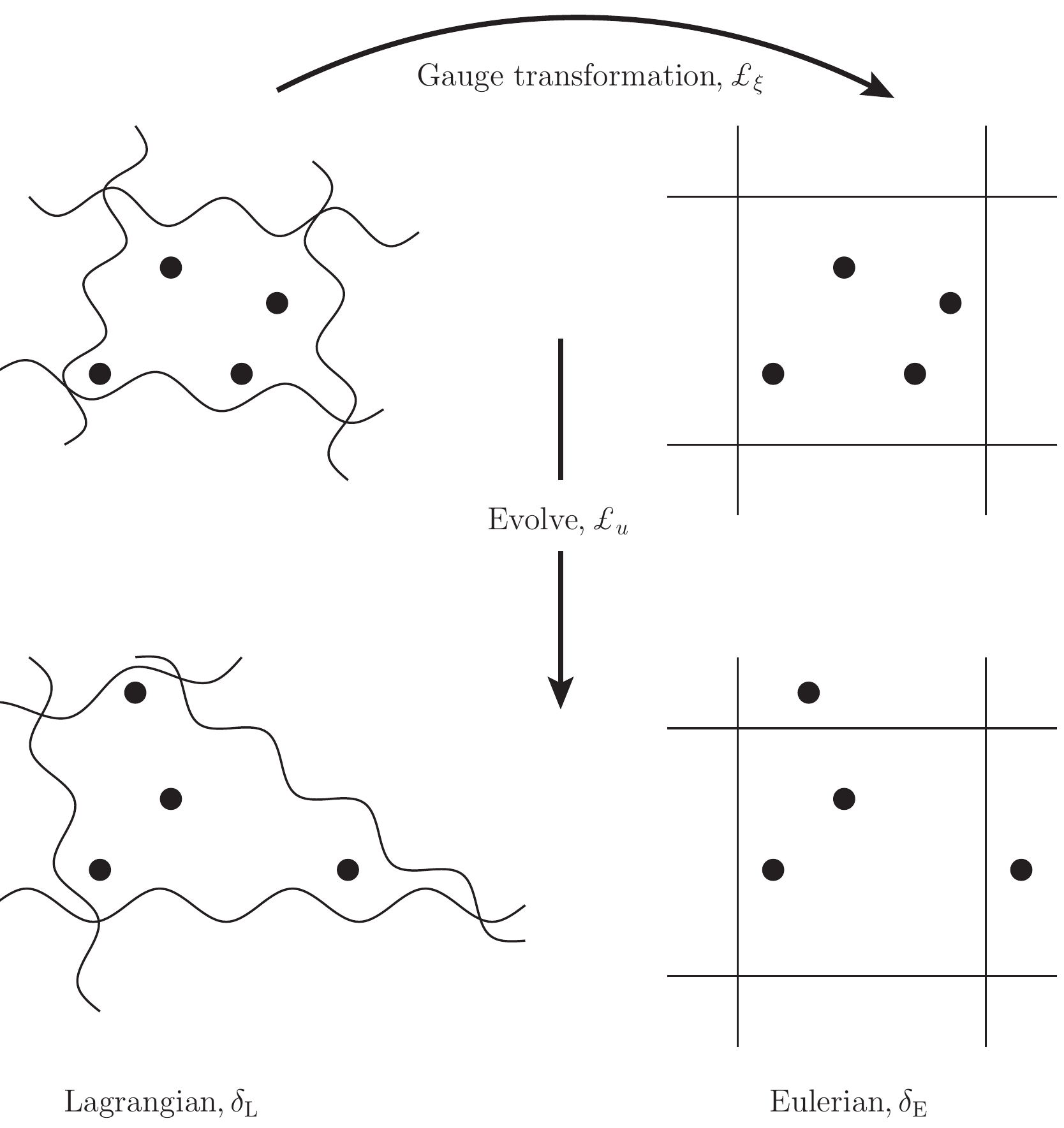}}
      \end{center}
\caption{Schematic view of the Eulerian and Lagrangian coordinate systems. The Lagrangian system can be said to be \textit{comoving}, and the Eulerian system as being \textit{fixed}. The Lagrangian system retains the density of a field, whereas the Eulerian system does not. This is schematically depicted by the ``grid square'' becoming deformed in the Lagrangian system on the left, to accommodate the movement of ``particles'' upon evolution in time. The grid square in the Eulerian system has remained fixed, meaning that the number of particles in a given square changes upon evolution. In cosmology we are perturbing against a \textit{fixed} background: the FRW background, however calculations are often easier to perform in a comoving system. This means that physical relevance is taken from equations perturbed according to a Eulerian scheme.} \label{fig:lagvsep}
\end{figure*}

Consider a quantity which is perturbed about some background value, $Q(t, \rbm{x}) = \bar{Q}(t) + \delta Q(t, \rbm{x})$. We can then employ two classes of coordinate system to follow the perturbation $\delta Q$ through evolution; time evolution can be thought of as Lie-dragging a quantity along a time-like vector, $u^{\mu}$,  to ``carve out'' the world-line of the perturbation, i.e. operating on a quantity with $\lied{u}$. The first is where the density of the perturbations remains fixed (i.e. the coordinate system evolves to   comove with the perturbations); this   is a \textit{Lagrangian} system. In the second, the coordinate system is fixed by some means (such as knowledge of the background geometry) and the density of the perturbations changes; this is an \textit{Eulerian} system. We write perturbations in the Lagrangian system as $\lp$ and perturbations in the Eulerian system as $\ep$. Evidently, a coordinate transformation can be used to transfer between the two systems, $x^{\mu}\rightarrow x^{\mu} + \xi^{\mu}$.  The Eulerian and Lagrangian variations are linked by  
\bea
\label{eq:sen-2.26-ep-lp}
\lp = \ep + \lied{\xi},
\eea
where $\lied{\xi}$ is the Lie derivative along the gauge field $\xi^{\mu}$.  This setup is schematically depicted in \fref{fig:lagvsep}.

\begin{figure*}[!t]
      \begin{center}
{\includegraphics[scale=0.7]{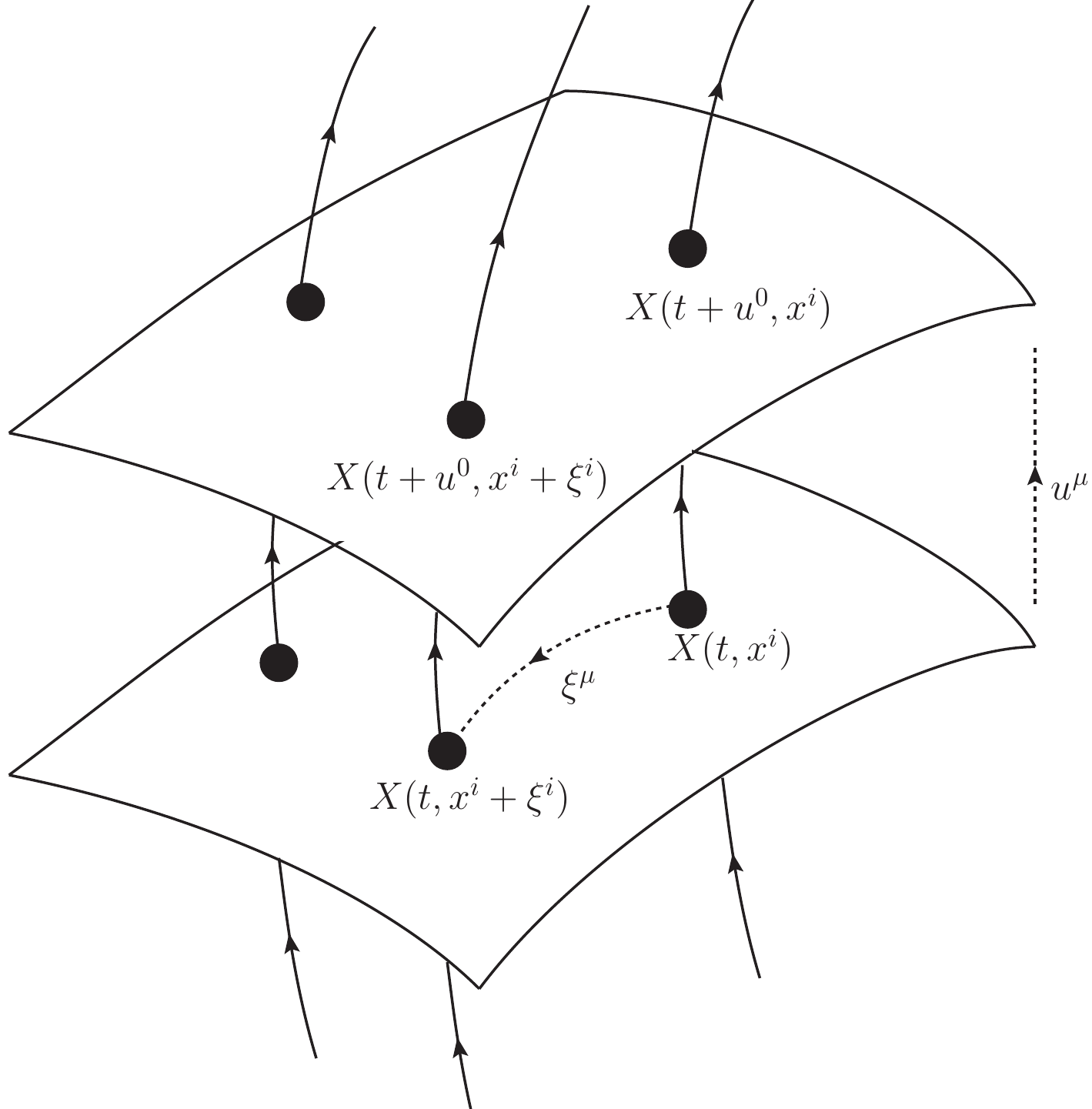}}
      \end{center}
\caption{Schematic view of the foliation and evolution, with three example world-lines drawn on, each piercing two 3D surfaces; $u^{\mu}$ is a time-like vector satisfying $u^{\mu}u_{\mu} = -1$. A quantity $X$ on a surface with spacetime location $(t, x^i)$ can be transformed into a quantity on the same surface  but at a different location by transforming the coordinate on the surface, $x^i \rightarrow x^i+\xi^i$. This is a diffeomorphism which drags one world-line into another. If the time coordinate is transformed $t \rightarrow t+u^0$ then the quantity is evaluated on a different 3d surface, but on the same world-line. Thus, if we were to have a transformation $x^{\mu} \rightarrow x^{\mu} + \xi^{\mu} + \chi u^{\mu}$, where $\chi$ is an arbitrary scalar field, the time-like part of $\xi^{\mu}$ is redundant. Hence, we are free to set $\xi^{\mu}u_{\mu}=0$, fixing the time-like part of the diffeomorphism field to be zero. So, we should have the interpretation that $\xi^{\mu}$ moves between world-lines and $u^{\mu}$ moves along world-lines.} \label{fig:foliation_evolve}
\end{figure*}

Without loss of generality we can set the gauge field $\xi^{\mu}$ and time-like vector $u_{\mu}$ to be mutually orthogonal,
\bea
\label{eq:xiu-ortho}
\xi^{\mu}u_{\mu} = 0.
\eea
This is because the  time-like transformations which the component $\xi^0$ could induce are world-line preserving, and are redundant when $u^{\mu}$ is present (which is   inherently a world-line preserving evolution). See \fref{fig:foliation_evolve} for a schematic view illustrating this point.

There is   an important question which arises: which perturbation scheme should we use to derive cosmologically relevant results,  i.e. which $\delta$ should we use: $\ep$ or $\lp$? In cosmological perturbation theory a quantity is perturbed from its value in a \textit{fixed} (or known) background (such as its value in an FRW background). Therefore, equations should be perturbed relative to a fixed background, and so we should employ the Eulerian scheme.  

The  equation of motion governing the metric perturbations is 
\bea
\label{eq:sec:eom-met-pert}
 \ep G_{\mu\nu} = 8 \pi G \ep T_{\mu\nu} + \ep U_{\mu\nu},
\eea
and the perturbed conservation law that should be solved is the one evaluated in a Eulerian system,
\bea
\ep \big( \nabla_{\mu}U^{\mu\nu}\big)=0,
\eea
which can be written as
\bea
\label{eq:sec:2.29-bi-ep}
\nabla_{\mu}\ep U^{\mu\nu} + \half \bigg[ U^{\mu\nu}g^{\alpha\beta} - U^{\alpha\beta}g^{\mu\nu} + 2 g^{\nu\beta}U^{\alpha\mu}\bigg]\nabla_{\mu}\ep g_{\alpha\beta}=0.
\eea

If the Lagrangian variation of the dark energy-momentum tensor is the quantity that is supplied, (i.e. $\lp U^{\mu\nu}$ is given), then one must be careful to use (\ref{eq:sen-2.26-ep-lp}),  to obtain the Eulerian perturbed quantity,
\bea
\label{eq:sec:2.29-ep-lp-lied-u}
\ep U^{\mu\nu} = \lp U^{\mu\nu} - \xi^{\alpha}\nabla_{\alpha}U^{\mu\nu} + 2U^{\alpha(\mu}\nabla_{\alpha}\xi^{\nu)}.
\eea
Furthermore, to obtain the components of the mixed Eulerian perturbed dark energy-momentum tensor, one must use
\bea
\ep {U^{\mu}}_{\nu} = g_{\alpha\nu}\ep U^{\mu\alpha} + U^{\mu\alpha}\ep g_{\nu\alpha}.
\eea
The Lagrangian and Eulerian perturbations of the metric are linked by
\bea
\label{eq:sec:3.25-lp-ep-g}
\ep g_{\mu\nu} =\lp g_{\mu\nu}  - 2 \nabla_{(\mu}\xi_{\nu)}.
\eea
For a vector field $A^{\mu}$ one finds that
\bea
 \ep A^{\mu} =\lp A^{\mu}-\xi^{\alpha} \nabla_{\alpha} A^{\mu} + A^{\alpha}\nabla_{\alpha}\xi^{\mu} .
\eea
As final explicit example, the Eulerian and Lagrangian variations of a scalar field $\phi$ are linked via
 \bea
\ep\phi = \lp\phi - \xi^{\mu}\nabla_{\mu}\phi.
\eea
An interesting lemma is that if $\nabla_{\mu}\phi \propto \dot{\phi}u_{\mu}$ then by (\ref{eq:xiu-ortho}) we find that the Eulerian and Lagrangian variations of a scalar field are identical, $\ep \phi = \lp \phi$. This means that  a diffeomorphism does not change the perturbations of the scalar field;   this is a   consequence of the background field being homogeneous.

%

\section{No extra fields: $\ld = \ld(g_{\mu\nu})$}
\label{sec:metriconly}
Our first and simplest example is where the dark sector does not contain any extra fields: only the metric is present, albeit in an arbitrary combination. This  class of theories contains   the cosmological constant and elastic dark energy \cite{PhysRevD.60.043505,   PhysRevD.76.023005}, and will also include more general theories that have not been previously considered. In this section we do not allow the dark sector to contain derivatives of the metric -- this is discussed in a subsequent section. One of the aims   is to build an intuition for understanding how to write down perturbative quantities and how to  decompose tensors which arise in the perturbative equations.

The Lagrangian density we will consider in this section is of the form
\bea
\ld = \ld (g_{\mu\nu}),
\eea
so that the second order Lagrangian is given   by 
\bea
\label{eq:met-only-lag}
\Diamond^2\ld = \frac{1}{8}\mathcal{W}^{\mu\nu\alpha\beta} \lp g_{\mu\nu} \lp g_{\alpha\beta}.
\eea
The rank-4 tensor $\mathcal{W}_{\mu\nu\alpha\beta} $ is only a function of background quantities, and is therefore manifestly gauge invariant.
We can use  (\ref{eq:Sec:per-u-prototype-formalismsection}) and (\ref{eq:met-only-lag})   to show that   the perturbations to the dark energy momentum tensor are given by
\bea
\label{eq:sec3.2-du-metonly}
\lp U^{\mu\nu} =-\half  \bigg\{\mathcal{W}^{\alpha\beta\mu\nu}  +  g^{\alpha\beta} U^{\mu\nu} \bigg\} \lp g_{\alpha\beta}.
\eea
 By inspecting (\ref{eq:met-only-lag}) it follows that the tensor $\mathcal{W}_{\alpha\beta\mu\nu}$ enjoys the following symmetries,
\bea
\label{eq:sec-w-ind-symm}
\mathcal{W}_{\alpha\beta\mu\nu}  = \mathcal{W}_{(\alpha\beta)(\mu\nu)}  = \mathcal{W}_{\mu\nu\alpha\beta}.
\eea
This shows us how to construct the   Lagrangian perturbations to the generalized gravitational field equations, under the assumption that the dark sector Lagrangian is a function of the metric only. Because it is the Lagrangian variation which appears above we must convert to Eulerian variations to obtain cosmologically relevant perturbations. By using (\ref{eq:sec:3.25-lp-ep-g}) and (\ref{eq:sec:2.29-ep-lp-lied-u}) in (\ref{eq:sec3.2-du-metonly}) we obtain
\bea
\label{eq:sec:3.5-ep-u-gonly}
\ep U^{\mu\nu} = -\half  \bigg\{\mathcal{W}^{\alpha\beta\mu\nu}  +  g^{\alpha\beta} U^{\mu\nu} \bigg\}\big( \ep g_{\alpha\beta} + 2\nabla_{(\alpha}\xi_{\beta)}\big) - \xi^{\alpha}\nabla_{\alpha}U^{\mu\nu} + 2U^{\alpha(\mu}\nabla_{\alpha}\xi^{\nu)}. \nonumber\\
\eea

To find the equation of motion of the vector field, $\xi^{\mu}$, we must compute the Eulerian perturbed Bianchi identity.  Substituting (\ref{eq:sec:3.5-ep-u-gonly}) into  (\ref{eq:sec:2.29-bi-ep})  we obtain
\bse
\label{eq:sec-3.4-pbi-metonly-totaleqn}
\bea
\label{eq:sec-3.4-pbi-metonly}
  2\bigg[L^{\mu\alpha\beta\nu}\bigg]\nabla_{\mu}\nabla_{\alpha}\xi_{\beta}+ 2\bigg[ \nabla_{\sigma}\mathcal{W}^{\sigma\nu\mu\alpha} \bigg]\nabla_{\mu}\xi_{\alpha} +2\bigg[\nabla_{\mu}\nabla_{\alpha}U^{\mu\nu}\bigg] \xi^{\alpha}  =\ep J^{\nu},
\eea
where, for convenience, we have defined
\bea
\label{eq:sec:l-3.5}
L^{\mu\alpha\beta\nu} \defn  \bigg[ \mathcal{W}^{\mu\nu\alpha\beta} + g^{\alpha\beta}U^{\mu\nu}- 2 U^{\alpha(\mu}g^{\nu)\beta}\bigg],
\eea
and where the perturbed source term, $\ep J^{\nu}$, is given by
\bea
\label{eq:sec:3.6-pert-Source}
\ep J^{\nu} \defn \bigg[  2 g^{\nu\beta}U^{\alpha\mu}  - U^{\alpha\beta}g^{\mu\nu} - \mathcal{W}^{\mu\nu\alpha\beta}  \bigg] \nabla_{\mu} \ep g_{\alpha\beta}- \bigg[ \nabla_{\mu}\mathcal{W}^{\mu\nu\alpha\beta}  \bigg]\ep g_{\alpha\beta} .
\eea
\ese 

 Here we observe that the metric perturbations $\ep g_{\mu\nu}$ and the diffeomorphism field $\xi^{\mu}$ are intimately linked: one cannot consistently set either to zero.
The equation (\ref{eq:sec-3.4-pbi-metonly-totaleqn}) is the   constraint equation for any parameters/functions that appear in a parameterization of the dark sector, under the rather general assumption that $\ld = \ld (g_{\mu\nu})$; the only freedom that remains is how to construct $\mathcal{W}_{\mu\nu\alpha\beta}$ out of background quantities.    In Section \ref{sec:cosm-perts} we will provide the components of the equation of motion for a perturbed FRW spacetime.
 
The only way to write the tensors $U_{\mu\nu}, \mathcal{W}_{\alpha\beta\mu\nu}$ with an isotropic (3+1) decomposition which respects the symmetries (\ref{eq:sec-w-ind-symm}) is
\bse
\label{eq:decomp-metriconly-proto}
\bea
U_{\mu\nu} = \rho u_{\mu}u_{\nu} + P \gamma_{\mu\nu},
\eea
\bea
\label{eq:decomp-w}
\mathcal{W}_{\mu\nu\alpha\beta} &=&  \AW  u_{\mu}u_{\nu}u_{\alpha}u_{\beta} + \BW  \bigg( \gamma_{\mu\nu}u_{\alpha}u_{\beta} + \gamma_{\alpha\beta}u_{\mu}u_{\nu}\bigg) \nonumber\\
&&+ 2\CW  \bigg( \gamma_{\mu(\alpha}u_{\beta)}u_{\nu} + \gamma_{\nu(\alpha}u_{\beta)}u_{\mu} \bigg)+\mathcal{E}_{\mu\nu\alpha\beta} ,
\eea
where $ \mathcal{E}_{\mu\nu\alpha\beta}$ respects the same symmetries as $ \mathcal{W}_{\mu\nu\alpha\beta}$,   satisfies $u^{\mu} \mathcal{E}_{\mu\nu\alpha\beta}=0$ (i.e.  $\mathcal{E}_{\mu\nu\alpha\beta}$ is entirely spatial) and is given by
\bea
\label{eq:e-gen-metonly-3.4c}
 \mathcal{E}_{\mu\nu\alpha\beta} = \DW   \gamma_{\mu\nu}\gamma_{\alpha\beta} + 2\EW  \gamma_{\mu(\alpha}\gamma_{\beta)\nu}.
 \eea
 \ese

A concrete example of   a theory which only contains the metric is the elastic dark energy theory \cite{PhysRevD.60.043505,   PhysRevD.76.023005} where one can find that the coefficients in terms of physical quantities such as energy density $\rho$, pressure $P$, bulk   $\beta$ and shear moduli $\mu$ are given by
\bse
\label{eq:sec:3.10-ede-coeffs}
\bea
\AW  = - \rho,\qquad \BW  = P,\qquad \CW  = -P,
\eea
\bea
 \DW  = \beta - P - \frac{2}{3}\mu,\qquad \EW  = \mu + P,
\eea
\ese
where the bulk modulus is defined via $\beta \defn (\rho + P) \frac{\dd P}{\dd\rho}$, and the pressure and shear modulus are functions of the density $P = P(\rho), \mu = \mu(\rho)$ (e.g. one way to choose these functional dependancies is with an ``equation of state'', $w$ and $\hat{\mu}$, so that $P = w\rho, \mu = \hat{\mu}\rho$).  
%

In this section we have identified that just five functions  are required to specify the perturbations in the dark sector when no extra fields are present. These five functions are
\bea
X = \bigg\{\AW ,\BW,\CW,\DW, \EW  \bigg\}
\eea
and each  function only depends on background quantities and are governed by the background evolution.

\section{Scalar fields: $\ld = \ld(g_{\mu\nu}, \phi, \nabla_{\mu}\phi)$}
\label{sec-darkfluids}
The second example   is when the dark sector contains an arbitrary combination of scalar field $\phi$, the  first derivative of the field $\nabla_{\mu}\phi$,  and the metric $g_{\mu\nu}$. This   encompasses scalar field theories such as quintessence and $k$-essence, but we could also encompass a range of other possible theories.

%

For a Lagrangian density given by
\bea
\ld = \ld (g_{\mu\nu}, \phi, \nabla_{\mu}\phi ),
\eea
the second order Lagrangian is given   by
\bea
\label{eq:sec:4.1-Diald-dakf}
\Diamond^2 \ld &=& \mathcal{A} ( \lp\phi)^2+\mathcal{B}^{\mu}  \lp\phi \nabla_{\mu} \lp\phi   +\half\mathcal{C}^{\mu\nu}\nabla_{\mu}\lp\phi \nabla_{\nu} \lp\phi  \nonumber\\
&&+ \frac{1}{4}\bigg[ \mathcal{Y}^{\alpha\mu\nu}   \nabla_{\alpha}\lp \phi\lp g_{\mu\nu}+ \mathcal{V}^{\mu\nu}\lp\phi\lp g_{\mu\nu}    +\half\mathcal{W}^{\mu\nu\alpha\beta}  \lp g_{\mu\nu} \lp g_{\alpha\beta}\bigg] . 
 \eea
The coefficients above comprise: one scalar $\mathcal{A}$, one vector $\mathcal{B}^{\mu}$, two rank-2 tensors $\mathcal{C}^{\mu\nu}, \mathcal{V}^{\mu\nu}$, one  rank-3 tensor, $\mathcal{Y}_{\alpha\mu\nu}$ and one rank-4 tensor $\mathcal{W}_{\mu\nu\alpha\beta}$, all of which are only functions of background quantities and are therefore gauge invariant. At first sight these are all independent quantities, but  we will show later that the conservation and Euler-Lagrange equations can be used to link the quantities. By inspecting (\ref{eq:sec:4.1-Diald-dakf}) these tensors enjoy the following symmetries,
\bse
\label{eq:sec-4.2-symm}
\bea
\mathcal{C}^{\mu\nu} = \mathcal{C}^{(\mu\nu)},\qquad \mathcal{Y}^{\alpha \mu\nu}=\mathcal{Y}^{\alpha (\mu\nu)},\qquad \mathcal{V}^{\mu\nu} = \mathcal{V}^{(\mu\nu)},
\eea
\bea
 \mathcal{W}^{\alpha\beta\mu\nu}  = \mathcal{W}^{(\alpha\beta)(\mu\nu)}  = \mathcal{W}^{\mu\nu\alpha\beta}.
\eea
\ese
 
 In what follows we will assume $\gamma^{\mu}_{\nu}\nabla_{\mu}\phi = 0$ (alternatively  this can be stated as $\xi^{\mu}\nabla_{\mu}\phi =0$), so that $\ep\phi = \lp\phi$. This is   the covariant statement that $\nabla_{\mu}\phi$ is entirely time-like,    while using the fact that the diffeomorphism is entirely space-like. Therefore, because the Eulerian and Lagrangian perturbations of a scalar field are identical we will not distinguish between them and we will write $\delta\phi \equiv \ep \phi = \lp \phi$.
 
The equation of motion of the perturbed scalar field,   $\delta\phi$, is given by the Euler-Lagrange equation (\ref{eq:sec-2.10-el-pert}). Using (\ref{eq:sec:4.1-Diald-dakf}) one finds
\bea
\label{eq:el-4.2-}
\mathcal{C}^{\mu\nu} \nabla_{\mu}\nabla_{\nu}\delta\phi + \big(\nabla_{\mu}\mathcal{C}^{\mu\nu}  \big)\nabla_{\nu}\delta\phi  +\big(\nabla_{\mu}\mathcal{B}^{\mu} -  2\mathcal{A}\big)\delta\phi  =\ep S,
\eea
where the ``perturbed source'' piece, $\ep S$, is given by
\bea
\ep S \defn \frac{1}{4}\bigg[ \big(\mathcal{V}^{\alpha\beta} - \nabla_{\mu}\mathcal{Y}^{\mu\alpha\beta}\big)\lp g_{\alpha\beta} - \mathcal{Y}^{\mu\alpha\beta}\nabla_{\mu}\lp g_{\alpha\beta}\bigg],
\eea
where $\lp g_{\alpha\beta} = \ep g_{\alpha\beta} + 2 \nabla_{(\alpha}\xi_{\beta)}$.
We note that $\mathcal{C}^{\mu\nu}$ plays the role of an ``effective metric'', due to its resemblance to the corresponding term in the perturbed Klein-Gordon equation, namely $g^{\mu\nu} \nabla_{\mu}\nabla_{\nu}\delta\phi$, and there is also an effective mass of the $\delta\phi$-field, $\qsubrm{M}{eff}^2 = \nabla_{\mu}\mathcal{B}^{\mu} - 2\mathcal{A}$.

The isotropic (3+1)  decomposition of the coefficients $\mathcal{A},\mathcal{B}^{\mu}, \mathcal{C}^{\mu\nu},\mathcal{V}_{\mu\nu}, \mathcal{Y}_{\alpha\mu\nu} $, whilst respecting the symmetries (\ref{eq:sec-4.2-symm}), is
\bse
\label{eq:sec-4.5-decomp-vy-df}
\bea
\mathcal{A} = A_{\mathcal{A}},
\eea
\bea
\mathcal{B}^{\mu} = A_{\mathcal{B}}u^{\mu} ,
\eea
\bea
\mathcal{C}_{\mu\nu} = A_{\mathcal{C}}u_{\mu}u_{\nu} + B_{\mathcal{C}}\gamma_{\mu\nu},
\eea
\bea
\mathcal{V}_{\mu\nu} = A_{\mathcal{V}}u_{\mu}u_{\nu} + B_{\mathcal{V}} \gamma_{\mu\nu},
\eea
\bea
\label{eq:sec-4.7-decomp-y}
\mathcal{Y}_{\alpha\mu\nu} = A_{\mathcal{Y}}u_{\alpha}u_{\mu}u_{\nu} + B_{\mathcal{Y}}u_{\alpha}\gamma_{\mu\nu} + 2C_{\mathcal{Y}} \gamma_{\alpha(\mu}u_{\nu)}.
\eea
\ese
The decompositions of $U_{\mu\nu}, \mathcal{W}_{\alpha\beta\mu\nu}$ are identical to those given in eq.(\ref{eq:decomp-metriconly-proto}).

Only the terms in (\ref{eq:sec:4.1-Diald-dakf}) which involve $\lp g_{\mu\nu}$ are relevant for writing down the perturbations to the dark energy-momentum tensor. We obtain
\bea
\label{eq:sec-decomp-darkfluids-preop}
\lp U^{\mu\nu} &=&  -\half \bigg\{   \mathcal{V}^{\mu\nu}\delta\phi +  \mathcal{Y}^{\alpha\mu\nu} \nabla_{\alpha}\delta \phi \bigg\}    -\half \bigg\{\mathcal{W}^{\alpha\beta\mu\nu}  +  g^{\alpha\beta} {} U^{\mu\nu} \bigg\}\lp g_{\alpha\beta}.
\eea
The Eulerian perturbed conservation law (\ref{eq:sec:2.29-bi-ep}) can be computed using (\ref{eq:sec-decomp-darkfluids-preop}).  We obtain
\bea
\label{eq:sec:4.6-pert-bian-iden}
 \mathcal{Y}^{\alpha\mu\nu}\nabla_{\mu}\nabla_{\alpha}\delta\phi + \big( \mathcal{V}^{\nu\alpha} + \nabla_{\mu}\mathcal{Y}^{\alpha\mu\nu}\big) \nabla_{\alpha}\delta\phi + \nabla_{\mu}\mathcal{V}^{\mu\nu} \delta\phi  = \ep J^{\nu} + 2 E^{\nu}  ,
 \eea
where   $\ep J^{\nu}$  is given by (\ref{eq:sec:3.6-pert-Source}), and $E^{\nu}$ represents the wave equation for $\xi^{\mu}$ and is given by
\bea
\label{eq:sec:e-pert-cont-shkfhdk}
E^{\nu} &\defn& -\big[L^{\mu\alpha\beta\nu}\big]\nabla_{\mu}\nabla_{\alpha}\xi_{\beta}  -\big[ \nabla_{\mu}\mathcal{W}^{\mu\nu\alpha\beta}\big]\nabla_{\alpha}\xi_{\beta} - \big[\nabla_{\mu}\nabla_{\alpha}U^{\mu\nu}\big]\xi^{\alpha}. 
\eea
Equation (\ref{eq:sec:4.6-pert-bian-iden}) is an evolution equation for the scalar field perturbation $\delta\phi$, sourced by the metric perturbations, $\ep g_{\mu\nu}$, and the vector field, $\xi^{\mu}$.  The scalar field perturbation sources the equation of motion for $\ep g_{\mu\nu}$,(\ref{eq:sec:eom-met-pert}), via the components $\ep {U^{\mu}}_{\nu}$. In general one cannot consistently solve the evolution equations for $\delta\phi$ independently from those for the vector field $\xi^{\mu}$; we will soon show  how these two  fields might decouple, but the decoupling only occurs in special cases.

The perturbed Euler-Lagrange equation (\ref{eq:el-4.2-}) and perturbed conservation law (\ref{eq:sec:4.6-pert-bian-iden}) are both evolution equations for $\delta\phi$, and both have apparently different coefficients, resulting in an over-determined system. 
 We can choose to  remove this apparent over-determination by ``forcing'' the time-like (i.e. scalar) part of the perturbed conservation law to be identical to the perturbed Euler-Lagrange equation. It is important to realize that the perturbed conservation law is a vector equation, and only one of its components can be set equal to the perturbed Euler-Lagrange equation; the other components of the vector equation  will introduce a set of constraint equations.  
 
When we contract the perturbed Bianchi identity with a time-like vector $\tau_{\mu} = \omega u_{\mu}$ (where $\nabla_{\mu}\omega = -u_{\mu}\dot{\omega}$), we can read off a set of conditions that link the coefficients appearing in the Euler-Lagrange equation (\ref{eq:el-4.2-}) and the perturbed Bianchi identity (\ref{eq:sec:4.6-pert-bian-iden}). Doing this we obtain the linking conditions
\bse
\label{eq:sec:5.7-consistency}
\bea
\label{eq:sec:4.7a-a-uy}
\mathcal{C}^{\mu\alpha} = \tau_{\nu} \mathcal{Y}^{\alpha\mu  \nu},
\eea
\bea
\label{eq:sec:5.7-b}
\nabla_{\mu}\mathcal{C}^{\mu\alpha} = \tau_{\nu} \big( \mathcal{V}^{\nu\alpha} + \nabla_{\mu}\mathcal{Y}^{\alpha\mu\nu}\big) ,
\eea
\bea
\label{eq:sec:5.7-c}
\nabla_{\mu}\mathcal{B}^{\mu} - 2\mathcal{A} = \tau_{\nu}\nabla_{\mu}\mathcal{V}^{\mu\nu},
\eea
\bea
\label{eq:sec:4.7d}
\ep S = \tau_{\nu}\bigg(\ep J^{\nu} + 2    E^{\nu} \bigg).
\eea
\ese
We see, therefore, that the coefficients $\{\mathcal{A}, \mathcal{B}^{\mu}, \mathcal{C}^{\mu\nu}\}$ and $\{\mathcal{V}^{\mu\nu}    ,\mathcal{Y}^{\alpha\mu\nu} , \mathcal{W}^{\alpha\beta\mu\nu}  \}$ that appear in $\Diamond^2\ld$ (\ref{eq:sec:4.1-Diald-dakf}) are not independent, which is now obvious from (\ref{eq:sec:5.7-consistency}).  
By differentiating (\ref{eq:sec:4.7a-a-uy}) and comparing with (\ref{eq:sec:5.7-b}) one   finds that
\bea
\label{eq:sec5.9-proper}
u_{\mu} u_{\nu}\mathcal{Y}^{\alpha\mu\nu}\dot{\omega} - (K_{\mu\nu}\mathcal{Y}^{\alpha\mu\nu}  - u_{\nu}\mathcal{V}^{\nu\alpha})\omega=0,
\eea
where $K_{\mu\nu} = \nabla_{\mu}u_{\nu}$ is the induced extrinsic curvature and  an overdot is used to denote differentiation along the time-like vector.

The (3+1) decomposition introduces some interesting structure and can be used to explicitly evaluate the linking conditions (\ref{eq:sec:5.7-consistency}). From (\ref{eq:sec:4.7a-a-uy}) we   find that 
\bea
A_{\mathcal{C}} = - \omega A_{\mathcal{Y}},\qquad B_{\mathcal{C}} = -  \omega C_{\mathcal{Y}}.
\eea
After combining (\ref{eq:sec:4.7a-a-uy}) and  (\ref{eq:sec:5.7-b}) to yield (\ref{eq:sec5.9-proper}) we find that
\bea
\dot{\omega}A_{\mathcal{Y}} - (A_{\mathcal{V}} + KB_{\mathcal{Y}})\omega=0.
\eea
In a similar fashion, it follows from (\ref{eq:sec:5.7-c})   that
\bea
\mathcal{A} = \half \bigg[ \dot{A}_{\mathcal{B}} + \omega \dot{A}_{\mathcal{V}} + K\big( A_{\mathcal{B}} + A_{\mathcal{V}} + B_{\mathcal{V}}\big) \bigg],
\eea
where $K = {K^{\mu}}_{\mu}$.

One can think of $\xi^{\mu}$ as being an ``artificial'' vector field whose role was to restore reparameterization invariance, and it would therefore be desirable to have a theory that does not require $\xi^{\mu}$ to be present and reparameterization invariance is manifest. We will derive conditions that the   tensors in the Lagrangian must satisfy in order for reparameterization invariance to be manifest. 

We  can rewrite the Lagrangian with the vector field $\xi^{\mu}$ explicitly present to show how the three   fields $\{\ep g_{\mu\nu},\vphi, \xi^{\mu}\}$ interact and how the parameters can be arranged so that they ultimately decouple. To ease our calculation we will write $h_{\mu\nu} \defn \ep g_{\mu\nu}$, We  use (\ref{eq:sec:3.25-lp-ep-g}) to replace $\lp g_{\mu\nu}$ with $h_{\mu\nu} + 2\nabla_{(\mu}\xi_{\nu)}$ in the Lagrangian (\ref{eq:sec:4.1-Diald-dakf}).  Rearranging, whilst keeping track of total derivatives yields
\bea
\label{eq:defn-lpg-epg-xi-2}
\sol &=&  \mathcal{A} ( \vphi)^2+\mathcal{B}^{\mu}  \vphi \nabla_{\mu}  \vphi   +\half\mathcal{C}^{\mu\nu}\nabla_{\mu} \vphi \nabla_{\nu} \vphi + \frac{1}{4}\bigg[   \half \mathcal{W}^{\mu\nu\alpha\beta} h_{\alpha\beta}\bigg]h_{\mu\nu} \nonumber\\
&&+ \frac{1}{4}\bigg[ \mathcal{V}^{\mu\nu}\vphi + \mathcal{Y}^{\alpha\mu\nu}\nabla_{\alpha}\vphi  \bigg]h_{\mu\nu} - \half \xi_{\nu}\bigg[ (\nabla_{\mu}\mathcal{W}^{\mu\nu\alpha\beta}) h_{\alpha\beta} + \mathcal{W}^{\mu\nu\alpha\beta}\nabla_{\mu}h_{\alpha\beta}   \bigg]\nonumber\\
&& - \half \xi_{\nu}\bigg[\mathcal{Y}^{\alpha\mu\nu}\nabla_{\mu}\nabla_{\alpha}\vphi + (\mathcal{V}^{\alpha\nu} + \nabla_{\beta}\mathcal{Y}^{\alpha\beta\nu})\nabla_{\alpha}\vphi + ( \nabla_{\mu}\mathcal{V}^{\mu\nu})\vphi \bigg]\nonumber\\
&& - \half \xi_{\nu}\bigg[   4 (\nabla_{\mu}\mathcal{W}^{\mu\nu\alpha\beta}) \nabla_{\alpha}\xi_{\beta} + 4 \mathcal{W}^{\mu\nu\alpha\beta} \nabla_{\mu}\nabla_{\alpha}\xi_{\beta} \bigg]\nonumber\\
&& + \half \nabla_{\alpha}\bigg[ \xi_{\beta}\big( \mathcal{Y}^{\mu\alpha\beta}\nabla_{\mu}\vphi + \mathcal{V}^{\alpha\beta}\vphi + \mathcal{W}^{\mu\nu\alpha\beta}h_{\mu\nu} + 2 \mathcal{W}^{\mu\nu\alpha\beta}\nabla_{\mu}\xi_{\nu} \big)\bigg].
\eea
To enable us to identify the ``free'' and ``interaction'' Lagrangians, we note that (\ref{eq:defn-lpg-epg-xi-2})  can be written schematically as
\bea
\sol &=& \subpsm{\ld}{\{2\}}{\rm A}[\vphi] + \subpsm{\ld}{\{2\}}{\rm B}[h_{\mu\nu}] +  \subpsm{\ld}{\{2\}}{\rm C}[\xi^{\alpha}] +  \subpsm{\ld}{\{2\}}{\rm D}[h_{\mu\nu},\vphi] \nonumber\\
&&+ \subpsm{\ld}{\{2\}}{\rm E} [h_{\mu\nu},\xi^{\alpha}] +  \subpsm{\ld}{\{2\}}{\rm F}[\vphi,\xi^{\alpha}] + \nabla_{\alpha}\mathcal{S}^{\alpha},
\eea
where $\subsm{\ld}{\{2\}}$ of a single field variable represents the self-interaction of that field and of two fields represents the interaction between the two fields.
The final line of (\ref{eq:defn-lpg-epg-xi-2}) is a pure surface term, and will not contribute to the dynamics, and thus does not require consideration in what we are about to discuss. However, if we note the definition of $\mathcal{S}^{\mu}$ and compare to the perturbed EMT (\ref{eq:sec-decomp-darkfluids-preop}), we find that $\mathcal{S}^{\mu} = - \xi_{\nu}( \lp U^{\mu\nu} + \half U^{\mu\nu}g^{\alpha\beta}\lp g_{\alpha\beta} )$.

Notice that the perturbed scalar field, $\vphi$, and vector field $\xi_{\mu}$ are   coupled in the Lagrangian, and only decouple when their interaction Lagrangian, $ \subpsm{\ld}{\{2\}}{\rm F}$, vanishes. This will  remove the \textit{direct} coupling but they may remain \textit{indirectly} coupled if the interaction Lagrangian for the perturbed metric and vector field remains non-zero (i.e. if $\subpsm{\ld}{\{2\}}{\rm E} \neq 0$), since the perturbed metric and scalar field will remain coupled, $\subpsm{\ld}{\{2\}}{\rm D}\neq 0$. So, the interaction Lagrangian between the vector field and perturbed scalar field vanishes, i.e. $ \subpsm{\ld}{\{2\}}{\rm F}=0$,  when
\bea
\label{eq:defn-lpg-epg-xi-3}
 \xi_{\nu}\bigg[\mathcal{Y}^{\alpha\mu\nu}\nabla_{\mu}\nabla_{\alpha}\vphi + (\mathcal{V}^{\alpha\nu} + \nabla_{\beta}\mathcal{Y}^{\alpha\beta\nu})\nabla_{\alpha}\vphi + ( \nabla_{\mu}\mathcal{V}^{\mu\nu})\vphi \bigg]=0.
\eea
For arbitrary values of the perturbed scalar field and vector field, this is satisfied by the covariant conditions
\bea
\label{eq:defn-lpg-epg-xi-3a}
\xi_{\nu}\mathcal{Y}^{\alpha\mu\nu} = 0,\qquad \xi_{\nu} (\mathcal{V}^{\alpha\nu} + \nabla_{\beta}\mathcal{Y}^{\alpha\beta\nu})=0,\qquad \xi_{\nu}\nabla_{\mu}\mathcal{V}^{\mu\nu}=0.
\eea

To find the decoupling conditions for the perturbed metric  we realize that because  $\ep (\nabla_{\mu}U^{\mu\nu})=0$,
\bea
\label{eq:sec:gen-decoup-cov-gen-ft-grhgdjfghkdfghfkd}
\xi_{\nu}\ep (\nabla_{\mu}U^{\mu\nu})=0
\eea
is an identity.  If we contract (\ref{eq:sec:4.6-pert-bian-iden}) with $\xi_{\mu}$ and use (\ref{eq:defn-lpg-epg-xi-3}) then
\bea
\label{eq:sec:decoupling-sf-fosft-one-two-hkdj-tot-3-da}
\xi_{\nu} \ep J^{\nu} + 2 \xi_{\nu}E^{\nu}=0,
\eea
where $\ep J^{\mu}$ and  $E^{\mu}$ are given respectively by (\ref{eq:sec:3.6-pert-Source}) and (\ref{eq:sec:e-pert-cont-shkfhdk}).
Inserting these definitions of $\ep J^{\nu}, E^{\nu}$    into (\ref{eq:sec:decoupling-sf-fosft-one-two-hkdj-tot-3-da}) yields
\bea
&& \xi_{\nu} (\mathcal{W}^{\mu\nu\alpha\beta} + U^{\alpha\beta} g^{\mu\nu} - 2 g^{\nu\beta}U^{\mu\alpha}) \nabla_{\mu}h_{\alpha\beta} + (\xi_{\nu}\nabla_{\mu}\mathcal{W}^{\mu\nu\alpha\beta})h_{\alpha\beta}\nonumber\\
&& +2\xi_{\nu} L^{\mu\alpha\beta\nu} \nabla_{\mu}\nabla_{\alpha}\xi_{\beta} + 2(\xi_{\nu} \nabla_{\mu}\mathcal{W}^{\mu\nu\alpha\beta})\nabla_{\alpha}\xi_{\beta} +2(\xi_{\nu}  \nabla_{\mu}\nabla_{\alpha}U^{\mu\nu})\xi^{\alpha}=0.  
\eea
For arbitrary values of $h_{\mu\nu}$, the decoupling of $\xi^{\mu}$ from $h_{\mu\nu}$ occurs when the coefficients of $\nabla_{\mu}h_{\alpha\beta}, h_{\alpha\beta}$ vanish, which occurs  when the covariant conditions
\bea
\label{eq:sec:decoup_metric_pre31-1}
\xi_{\nu}( \mathcal{W}^{\mu\nu\alpha\beta} + U^{\alpha\beta} g^{\mu\nu} - 2 g^{\nu\beta}U^{\mu\alpha})=0,\qquad
\xi_{\nu}\nabla_{\mu}\mathcal{W}^{\mu\nu\alpha\beta} = 0
\eea
are satisfied. 

Inserting the (3+1) decomposition into (\ref{eq:defn-lpg-epg-xi-3a}) and (\ref{eq:sec:decoup_metric_pre31-1})  allows us to evaluate the decoupling conditions. This  yields
\bse
\label{eq:sec:decoup_sf}
\bea
\xi_{\nu} \mathcal{Y}^{\alpha\mu\nu} = B_{\mathcal{Y}}u^{\alpha}\xi^{\mu}+C_{\mathcal{Y}}u^{\mu}\xi^{\alpha},
\eea
\bea
\xi_{\nu}( \mathcal{V}^{\alpha\nu} + \nabla_{\beta}\mathcal{Y}^{\alpha\beta\nu} )=  \bigg[ \dot{C}_{\mathcal{Y}} +  {C}_{\mathcal{Y}} K + B_{\mathcal{V}} \bigg] \xi^{\alpha}+ \bigg[  {B}_{\mathcal{Y}} + {C}_{\mathcal{Y}}  \bigg]\xi_{\nu}K^{\alpha\nu},
\eea
\bea
\label{eq:sec:decoupling-sf-fosft-one-two-hkdj-a}
\xi_{\nu} \nabla_{\mu}\mathcal{V}^{\mu\nu} = \bigg[ \dot{A}_{\mathcal{V}} + (A_{\mathcal{V}} + B_{\mathcal{V}})K\bigg] \xi_{\nu} u^{\nu}=0.
\eea
\ese
\bse
\label{eq:sec:decoup_met}
\bea
\label{eq:sec:decoup-h-boiiiii}
(\BW+\rho) \xi^{\mu}u^{\alpha}u^{\beta}  + 2( \CW -\rho)\xi^{(\alpha}u^{\beta)}u^{\mu}    +( \DW+P) \xi^{\mu}\gamma^{\alpha\beta}  + 2(\EW-P) \xi^{(\alpha}\gamma^{\beta)\mu}=0,\nonumber\\
\eea
\bea
\label{eq:sec:decoup-h-boiiiii-2}
2\bigg[ \dot{C}_{\mathcal{W}} + K(\CW + \EW)\bigg] u^{(\alpha}\xi^{\beta)}  + 2\bigg[ \BW + \CW + \DW + \EW\bigg] \xi^{\mu}{K^{(\alpha}}_{\mu}u^{\beta)}=0.\nonumber\\
\eea
\ese
Note that (\ref{eq:sec:decoupling-sf-fosft-one-two-hkdj-a}) gives us no information since $u^{\mu}\xi_{\mu}=0$.
Hence, we conclude  that the decoupling conditions (\ref{eq:sec:decoup_sf}, \ref{eq:sec:decoup_met}) are satisfied by the parameter choices
\bse
\label{eq:sec:decoup-h-boiiiii-3}
\bea
\label{eq:sec:decoupling-sf-fosft-one-two-hkdj}
\dot{C}_{\mathcal{Y}} +  {C}_{\mathcal{Y}} K + B_{\mathcal{V}}=0,\qquad  {B}_{\mathcal{Y}} =- {C}_{\mathcal{Y}}  .
\eea
\bea
\label{eq:sec:decoup-h-boiiiii-33}
\BW = - \rho,\qquad \CW = \rho,\qquad \DW = -P,\qquad \EW = P,
\eea
 \bea
\label{eq:sec:decoup-h-boiiiii-334}
\dot{C}_{\mathcal{W}} + K(\CW + \EW)=0,\qquad \BW + \CW + \DW + \EW=0.
\eea
\ese
When (\ref{eq:sec:decoup-h-boiiiii-33}) is used  the first condition of (\ref{eq:sec:decoup-h-boiiiii-334})   becomes
\bea
\dot{\rho} + K(\rho+P)=0,
\eea
and the second is satisfied identically.


The process of identifying the time-like part of the perturbed    conservation law with the Euler-Lagrange equation has reduced the number of functions required to specify $\Diamond^2\ld$ from $14\rightarrow 11$. The eleven   functions  are
\bea
\label{eq:sec:params-decomp-fosft}
\bigg\{A_{\mathcal{B}},\AW  , \BW,\CW,\DW, \EW  , A_{\mathcal{V}}, B_{\mathcal{V}}, A_{\mathcal{Y}}, B_{\mathcal{Y}}, C_{\mathcal{Y}}\bigg\}  ,
\eea
as well as the energy density $\rho$ and pressure $P$ of the dark sector ``fluid''.   Imposing reparameterization invariance as well, these eleven functions reduce to just five:
\bea
\bigg\{ A_{\mathcal{B}}, \AW, A_{\mathcal{V}},A_{\mathcal{Y}},C_{\mathcal{Y}}\bigg\}.
\eea
Later on we will show that $A_{\mathcal{B}}$ does not affect the cosmological dynamics, and $\AW$ becomes irrelevant in the synchronous gauge. This means that there are just three free functions left to completely specify the dark sector perturbations  for a reparameterization-invaraiant scalar field theory.

In Appendix \ref{appendic:darkfluids} we provide  the explicit calculation for computing $\Diamond^2\ld$ and $\lp U^{\mu\nu}$ in  a kinetic scalar field theory $\ld = \ld(\kin, \phi)$, where $\mathcal{X} = -\half g^{\mu\nu} \nabla_{\mu}\phi \nabla_{\nu}\phi$ is the kinetic term of a scalar field.   In \tref{table:summaryparameters} we give a summary of the functions that appear in the decomposition of $\lp U^{\mu\nu}$ for  some explicit scalar field theories; in these examples it is natural to see that upon specifying a scalar field theory the time evolution of the various functions is set.  For a canonical scalar field theory, $\ld = \kin - V(\phi)$, the functions   are given by
\bse
\label{eq:5.13-canonical-sft}
\bea
\rho =   \half \dot{\phi}^2 + V,\qquad  P =   \half\dot{\phi}^2 - V,
\eea
\bea
A_{\mathcal{V}} = - B_{\mathcal{V}} =- 2 V',
\eea
\bea
A_{\mathcal{Y}} = B_{\mathcal{Y}} =  -C_{\mathcal{Y}} = -2 \dot{\phi},
\eea
\bea
\AW =-(2\rho +P),\qquad \BW = -\CW = - \rho,\qquad \DW = - \EW = -P,
\eea
\ese
where an overdot is understood to denote differentiation with respect to time and $V' = \dd V/\dd\phi$.  Using (\ref{eq:5.13-canonical-sft}) and taking $\omega = 1/\dot{\phi}$ it transpires that  (\ref{eq:sec5.9-proper}) is the Klein-Gordon equation.

\begin{table}[!t]\footnotesize
\begin{center}
\begin{tabular}{|c||c||c||c|| c|}
\hline
Function & (a) EDE & (b) $\ld = \ld(\phi, \kin) $& (c) $\ld = F(  \kin)$& (d) $\ld = \kin - V(\phi)$\\\hline\hline
$A_{\mathcal{V}}$ &  $ 0$ & $-2(\ld_{,\kin\phi}\dot{\phi}^2-\ld_{,\phi})$ & 0 & $-2V'$\\
$B_{\mathcal{V}}$ & $0$ & $ -2\ld_{,\phi} $ & 0 & $2V'$\\
\hline
$A_{\mathcal{Y}}$ & $0$ &$ -2(\ld_{,\kin\kin}\dot{\phi}^3+\ld_{,\kin}\dot{\phi}) $& $-2(F''\dot{\phi}^2+F'\dot{\phi})$ & $-2\dot{\phi}$\\
$B_{\mathcal{Y}}$ & $0$& $ -2\ld_{,\kin}\dot{\phi} $ & $-2F'\dot{\phi}$ & $-2\dot{\phi}$\\
$C_{\mathcal{Y}}$ & $0$ & $2\ld_{,\kin}\dot{\phi} $ & $2F'\dot{\phi}$ &$2\dot{\phi}$\\
\hline
$\AW$ & $-\rho$ & $-(\ld_{,\kin\kin}\dot{\phi}^4+2\rho+P) $ & $-(F''\dot{\phi}^4 +2\rho +P)$ &$-(2\rho+P)$\\
$\BW$ & $P$ & $-\rho$& $-\rho$&$-\rho$\\
$\CW$ & $-P$& $\rho$& $\rho$& $\rho$\\
$\DW$ & $\beta-P-\frac{2}{3}\mu$& $-P$& $-P$ &$-P$\\
$\EW$ & $\mu+P$& $P$ &$P$&$P$\\
\hline
\end{tabular}\caption{Collection of the functions in the decomposition of $\lp U^{\mu\nu}$. The theories we have presented are: (a) elastic dark energy, (b) generic kinetic scalar field theory, (c) $k$-essence and (d) canonical scalar field theory. It is interesting to realize that the theories with $\ld = \ld(g_{\mu\nu})$ are subsets of theories with $\ld = \ld(g_{\mu\nu},\phi, \nabla_{\mu}\phi)$. Comma denotes partial differentiation (e.g. $\ld_{,\phi} = \partial\ld/\partial\phi$), prime denotes differentiation with respect the functions single argument: $F' = \dd F/\dd \kin$, $V' = \dd V/\dd\phi$ and an overdot denotes differentiation with respect to time; for conformal time coefficients one should replace $\dot{\phi} \rightarrow \dot{\phi}/a$.  The free function $A_{\mathcal{B}}$ does not appear in the decomposition of $\lp U^{\mu\nu}$, but for the sake of completeness its value in an $\ld = \ld(\phi, \kin)$-theory is $A_{\mathcal{B}} = -\ld_{,\phi\kin}\dot{\phi}$ and in the cases (c, d), $A_{\mathcal{B}}=0$.}\label{table:summaryparameters}
\end{center}
\end{table}

\section{Cosmological perturbations}
\label{sec:cosm-perts}
In this section we provide explicit expressions for the components of $\ep {U^{\mu}}_{\nu}$ and the perturbed conservation equation specialized to the case of an FRW background.  We will pay special attention to the   scalar field theory, where we will show how the vector field $\xi^{\mu}$ decouples from the equation of motion for $\delta\phi$.  

 We will perturb the line element about a conformally flat FRW background, and write 
\bea
\dd s^2 = a^2(\tau)\bigg[ - (1 - 2\Phi) \dd \tau^2 + 2N_i \dd x^i \dd \tau + (\delta_{ij} + h_{ij}) \dd x^i \dd x^j \bigg].
\eea
This means that we are setting the components of the Eulerian perturbed metric to
 \bea
\ep g_{00} = 2a^2(\tau)\Phi(\tau, \rbm{x}),\qquad \ep g_{0i} = a^2(\tau)N_i(\tau, \rbm{x}),\qquad \ep g_{ij} = a^2(\tau) h_{ij}(\tau, \rbm{x}).
\eea
 The time-like vector is given by $u_{\mu} = a(\tau)(-1,0,0,0)$, and we   set $\xi^{\mu}u_{\mu}=0$.  All functions (\ref{eq:sec:params-decomp-fosft}) are only functions of time. The background conservation equation $\nabla_{\mu}U^{\mu\nu}=0$ becomes
\bea
\dot{\rho} + 3 \hct (\rho +P)=0,
\eea
where an overdot denotes derivative with respect to conformal time $\tau$ and $\hct$ is the conformal time Hubble parameter.  The components of $\ep {U^{\mu}}_{\nu}$ for the theory with field content $\ld = \ld(g_{\mu\nu}, \phi, \nabla_{\mu}\phi)$, (\ref{eq:sec-decomp-darkfluids-preop}),  are given by
\bse
\label{eq:sec:8.4-epU-genmetrci}
\bea
\ep {U^0}_0 = (\rho  + \BW)\bigg( \partial_k \xi^k + \half h\bigg) + (\rho  + \AW)\Phi + \half \bigg( A_{\mathcal{V}} \delta\phi + \frac{1}{a}A_{\mathcal{Y}} \dot{\delta\phi}\bigg),
\eea
\bea
\ep {U^i}_0 = (\CW - \rho ) \dot{\xi}^i + (P  + \CW)N^i + \frac{1}{2a} C_{\mathcal{Y}} \partial^i\delta\phi,
\eea
\bea
\ep {U^0}_i = \big( \rho  - \CW\big) \bigg( \dot{\xi}_i + N_i\bigg) - \frac{1}{2a}C_{\mathcal{Y}}\partial_i\delta\phi,
\eea
\bea
\ep {U^i}_j &=& - \bigg\{(\DW + P )\bigg( \partial_k\xi^k + \half h\bigg)  + (\BW-P )\Phi + \half \bigg( B_{\mathcal{V}}\delta\phi + \frac{1}{a}B_{\mathcal{Y}} \dot{\delta\phi}  \bigg) \bigg\}{\delta^i}_j\nonumber\\
&& + (P  - \EW)\bigg({h^i}_j + \partial^i\xi_j + \partial_j\xi^i\bigg).
\eea
\ese 
These are the sources to the equations governing the evolution of the metric perturbations, and can be used to obtain the components of $\ep {U^{\mu}}_{\nu}$ in the conformal Newtonian and synchronous gauges.

We will now work in the synchronous gauge  (by setting $\Phi = N_i = 0$), and we will study the more general theory $\ld = \ld(g_{\mu\nu}, \phi, \nabla_{\mu}\phi)$, which will trivially encompass the no-extra-fields case. The components $\ep {U^{\mu}}_{\nu}$ become
\bse
\label{eq:sec:dsft-deu}
\bea
\ep {U^0}_0 = (\rho  + \BW)\bigg( \partial_k \xi^k + \half h\bigg)   + \half \bigg( A_{\mathcal{V}} \delta\phi + \frac{1}{a}A_{\mathcal{Y}} \dot{\delta\phi}\bigg),
\eea
\bea
\ep {U^i}_0 = (\CW - \rho ) \dot{\xi}^i   + \frac{1}{2a} C_{\mathcal{Y}} \partial^i\delta\phi,
\eea
\bea
\ep {U^i}_j &=& - \bigg\{(\DW + P )\bigg( \partial_k\xi^k + \half h\bigg)   + \half \bigg( B_{\mathcal{V}}\delta\phi + \frac{1}{a}B_{\mathcal{Y}} \dot{\delta\phi}  \bigg) \bigg\}{\delta^i}_j\nonumber\\
&& + (P  - \EW)\bigg({h^i}_j + \partial^i\xi_j + \partial_j\xi^i\bigg).
\eea
\ese
The components of the Eulerian perturbed conservation law $\ep (\nabla_{\mu}U^{\mu\nu})=0$, (\ref{eq:sec:4.6-pert-bian-iden}),  has a ``scalar'' $\nu=0$ component and a ``vector'' $\nu=i$ component. The     $\nu = 0$ component of the perturbed conservation law (\ref{eq:sec:4.6-pert-bian-iden}) yields
\bse
\label{eq:8.11-wow}
 \bea
 \label{eq:8.11-wow-a}
&&A_{\mathcal{Y}} \ddot{\delta\phi } +C_{\mathcal{Y}}\nabla^2\delta\phi  +   \bigg[  \dot{A}_{\mathcal{Y}} +aA_{\mathcal{V}}+(2A_{\mathcal{Y}}+3B_{\mathcal{Y}} )\hct \bigg]\dot{\delta\phi}  + a\bigg[\dot{A}_{\mathcal{V}} +3\hct(A_{\mathcal{V}} +B_{\mathcal{V}})\bigg]\delta\phi\nonumber \\
&&\qquad\qquad\qquad\qquad\qquad=- a\bigg[\dot{B}_{\mathcal{W}} + \hct(3\BW+3\DW + 2 \EW  -2P) \bigg](h+2\partial_i\xi^i)\nonumber\\
&&\qquad\qquad\qquad\qquad\qquad\qquad +a\bigg[P-\BW\bigg]\dot{h} - 2a\bigg[\CW+\BW\bigg]\partial_i\dot{\xi}^i,
\eea
and the   $\nu = i$ component yields
\bea
\label{eq:8.12-b}
&&\bigg[\rho-\CW\bigg](\ddot{\xi}^i + \hct \dot{\xi}^i) - \bigg[\dot{C}_{\mathcal{W}} + 3 \hct(\CW+P)\bigg] \dot{\xi}^i\nonumber\\
&&  \qquad\qquad\qquad- \bigg[\DW+\EW\bigg]\partial^i\partial_k\xi^k + \bigg[P-\EW\bigg]\partial_k\partial^k\xi^i  \nonumber\\
&&\qquad\qquad\qquad\qquad\qquad=\frac{1}{2a}\bigg[ B_{\mathcal{Y}}+C_{\mathcal{Y}}\bigg]\partial^i\dot{\delta\phi}   +\frac{1}{2a}  \bigg[\dot{C}_{\mathcal{Y}}+ 3C_{\mathcal{Y}}\hct +aB_{\mathcal{V}} \bigg]\partial^i\delta\phi  \nonumber\\
&&\qquad\qquad\qquad\qquad\qquad\qquad-  \bigg[P-\EW\bigg]\partial^j{h^i}_j +\half \bigg[\DW+P\bigg]\partial^ih.
\eea
\ese
We   observe that the ``scalar'' piece (\ref{eq:8.11-wow-a}) of the perturbed conservation law represents the evolution equation for the perturbed scalar field sourced by   metric perturbations and the vector field, and the ``vector'' piece (\ref{eq:8.12-b}) constitutes an evolution equation for the vector field, sourced by the perturbed scalar field and metric perturbations.   Notice that nine functions are required to be specified to be able to write down the components $\ep {U^{\mu}}_{\nu}$ and the perturbed conservation law: $A_{\mathcal{V}}, B_{\mathcal{V}}$, $A_{\mathcal{Y}}, B_{\mathcal{Y}}, C_{\mathcal{Y}}, \BW, \CW, \DW, \EW$ (note that $ {A}_{\mathcal{B}}, {A}_{\mathcal{W}}$ do not enter into these quantities).

We will now study the conditions under which $\xi^{\mu}$ and $\delta\phi$ decouple; this will represent a simpler subset of   theories, and will provide us with  another understanding how   the     decoupling conditions come about. It is useful to  write the perturbed conservation equation (\ref{eq:8.11-wow}) as
\bse
\label{eq:sec:decoup-dphi-xi}
\bea
\mathcal{C}_1 \ddot{\delta\phi } +\mathcal{C}_2\nabla^2\delta\phi  +   \mathcal{C}_3\dot{\delta\phi}  + \mathcal{C}_4\delta\phi =\mathcal{D}_1  (h+2\partial_i\xi^i) 
 +\mathcal{D}_2\dot{h} +\mathcal{D}_3\partial_i\dot{\xi}^i,
\eea
\bea
\mathcal{F}_1\ddot{\xi}^i + \mathcal{F}_2 \dot{\xi}^i + \mathcal{F}_3 \partial^i\partial_k\xi^j + \mathcal{F}_4\partial_k\partial^k\xi^i  = \mathcal{G}_1 \partial^ih + \mathcal{G}_2 \partial^j{h^i}_j + \mathcal{G}_3\partial^i\delta\phi + \mathcal{G}_4\partial^i\dot{\delta\phi},
\eea
\ese
where the sets of coefficients $\{\mathcal{C}_{(\rm A)}, \mathcal{D}_{(\rm A)}, \mathcal{F}_{(\rm A)}, \mathcal{G}_{(\rm A)}\}$ can be read off from (\ref{eq:8.11-wow}). The $\xi^{\mu}$ and $\delta\phi$   decouple when all common terms in (\ref{eq:sec:decoup-dphi-xi}) vanish. This yields the conditions $\mathcal{D}_1 = \mathcal{D}_3=0$ and   $\mathcal{G}_1 = \mathcal{G}_2 = \mathcal{G}_3 = \mathcal{G}_4 = 0$. The former decoupling condition yields
\bse
\label{eq:sec:sft-decoup}
\bea
\label{eq:sec:e0-decouple-b}
\dot{B}_{\mathcal{W}}+\hct(3\BW +3\DW + 2 \EW -2P     )=0
\eea
\bea
\label{eq:sec:e0-decouple}
C_{\mathcal{W}} = - B_{\mathcal{W}},
\eea
and the latter decoupling condition yields
\bea
\label{eq:sec:e0-decouple-4}
\DW = -P,\qquad \EW = P,\qquad 
B_{\mathcal{Y}}=-C_{\mathcal{Y}}.
\eea
\bea
\label{eq:sec:e0-decouple-c}
\dot{C}_{\mathcal{Y}}     +3 C_{\mathcal{Y}}\hct +a B_{\mathcal{V}} =0.
\eea
We also require that 
\bea
\label{eq:sec:e0-decouple-fghdj}
\BW = - \rho
\eea
\ese 
 for decoupling to occur in the $\ep {U^{\mu}}_{\nu}$. Combining (\ref{eq:sec:e0-decouple-b}, \ref{eq:sec:e0-decouple-4}, \ref{eq:sec:e0-decouple-fghdj}) yields the conservation equation: $\dot{\rho} + 3 \hct(\rho + P)=0$.  These conditions are compatible with those we derived covariantly, (\ref{eq:sec:decoup-h-boiiiii-3}).

Applying the decoupling conditions (\ref{eq:sec:sft-decoup}) to the components $\ep {U^{\mu}}_{\nu}$ (\ref{eq:sec:dsft-deu}) we obtain
\bse
\bea
\ep {U^0}_0 = \half \bigg( A_{\mathcal{V}} \delta\phi + \frac{1}{a}A_{\mathcal{Y}}\dot{\delta\phi}\bigg),\qquad
\ep {U^i}_0 = \frac{1}{2a} C_{\mathcal{Y}} \partial^i\delta\phi,
\eea
\bea
\ep {U^i}_j = \frac{1}{2a}\bigg( (\dot{C}_{\mathcal{Y}} + 3 \hct C_{\mathcal{Y}})\delta\phi + C_{\mathcal{Y}}\dot{\delta\phi}\bigg),
\eea
\ese
and to the perturbed conservation equation (\ref{eq:8.11-wow-a}) we obtain 
\bea
&&A_{\mathcal{Y}}\ddot{\delta\phi} + C_{\mathcal{Y}}\nabla^2\delta\phi +\bigg[  \dot{A}_{\mathcal{Y}} +aA_{\mathcal{V}}+(2A_{\mathcal{Y}}-3C_{\mathcal{Y}} )\hct \bigg]\dot{\delta\phi}\nonumber\\
&&\qquad + \bigg[ a\dot{A}_{\mathcal{V}} + 3a \hct A_{\mathcal{V}} - 3 \hct (\dot{C}_{\mathcal{Y}} + 3 \hct C_{\mathcal{Y}})\bigg]\delta\phi = a (\rho+P)\dot{h}.
\eea
We now observe that only three functions are required to be specified: $A_{\mathcal{V}}, A_{\mathcal{Y}}, C_{\mathcal{Y}}$. When the decoupling conditions are satisfied, one can consistently set $\xi^{\mu}=0$   and reparameterization invariance is enforced. The theory is now equivalent to the theory studied in \cite{Creminelli:2008wc}, where    reparameterization  invariance was imposed from the implicitly. 

\subsection{No extra fields}
For the theory with no extra fields, we can obtain the components of $\ep {U^{\mu}}_{\nu}$ and those of the perturbed conservation law by ignoring all terms with $\delta\phi$ in (\ref{eq:sec:dsft-deu}) and (\ref{eq:8.11-wow}) respectively.
The components  $\ep {U^{\mu}}_{\nu}$ become
\bse
\label{eq:sec:no-eft-deu}
\bea
\label{eq:sec:6.2a-deltau00}
\ep {U^0}_0 = (\rho  + \BW )\bigg( \partial_k \xi^k + \half h\bigg),
\eea
\bea
\ep {U^i}_0 = - (\rho  - \CW )\dot{\xi}^i,
\eea
\bea
\ep {U^i}_j = - (\DW  + P )\bigg(\partial_k \xi^k + \half h\bigg){\delta^i}_j + (P  - \EW )({h^i}_j + \partial^i\xi_j + \partial_j \xi^i).
\eea
\ese
The $\nu = i$ component of the perturbed conservation law becomes
\bea
\label{eq:sec:6.4fd}
&&\bigg[\rho -\CW  \bigg] \big(\ddot{\xi}^i + \hct \dot{\xi}^i\big)  - \bigg[ \dot{C}_{\mathcal{W}} +3 \hct (\CW  +P )  \bigg]\dot{\xi}^i- \bigg[\DW  +\EW \bigg] \partial^i\partial_k\xi^k   \nonumber\\   
&&\qquad \qquad +\bigg[P -\EW \bigg]\partial_k\partial^k\xi^i   = \frac{1}{2}(\DW +P )\partial^ih + (\EW-P )\partial^k{h^i}_k,
\eea
and  the $\nu=0$ component of the perturbed conservation law yields
\bea
\label{eq:sec:6.5}
&& \bigg[\dot{B}_{\mathcal{W}}+\hct\left(3\BW +3\DW+2\EW -2P     \right) \bigg]\bigg( \partial_i\xi^i+\half h\bigg)\nonumber\\
&&\qquad \qquad+ \bigg[\CW +\BW \bigg]\partial_i\dot{\xi}^i  =   \half  \bigg[\BW-P \bigg]  \dot{h}. 
\eea
 
We can use (\ref{eq:sec:6.5}) to obtain a set of conditions that enforces the constraint (\ref{eq:sec:6.5}) on the $\nu=i$ component of the perturbed conservation equation. We  set the coefficients of $(\partial_i\xi^i+\half h), \partial_i\dot{\xi}^i, \dot{h}$ to zero and obtain
\bse
\label{eq:sec:const-xi-1}
\bea
\BW =-\CW =P,
\eea
\bea
\dot{P}+\hct\big(P  + 3\DW  + 2 \EW  \big)=0.
\eea
\ese
Applying these   conditions to the components of $\ep {U^{\mu}}_{\nu}$ (\ref{eq:sec:no-eft-deu}) we find
\bse
\bea
\ep {U^0}_0 = (\rho  + P )\bigg( \partial_k \xi^k + \half h\bigg),
\eea
\bea
\ep {U^i}_0 = - (\rho  +P )\dot{\xi}^i,
\eea
\bea
\ep {U^i}_j = - (\DW  + P )\bigg(\partial_k \xi^k + \half h\bigg)\delta^i_j + (P  - \EW )({h^i}_j + \partial^i\xi_j + \partial_j \xi^i),
\eea
\ese
and the equation of motion (\ref{eq:sec:6.4fd}) becomes
 \bea
 \label{eq:sec:eqn-becomes-ede-after-choice}
 &&\bigg[\rho +P  \bigg] \ddot{\xi}^i   +\hct \bigg[ \rho  - 3 \DW - 2 \EW   \bigg]\dot{\xi}^i   -  (\DW  +\EW ) \partial^i\partial_k\xi^k - \frac{1}{2}(\DW +P )\partial^ih \nonumber\\   
&&\qquad \qquad+(P -\EW )\bigg[\partial_k\partial^k\xi^i +  \partial^k{h^i}_k   \bigg]   =0. 
\eea
Hence, we see that after applying the   conditions (\ref{eq:sec:const-xi-1}) there are only two free coefficients which describe perturbations in the dark sector: $\DW$ and $\EW$.
 Making the choice which defines elastic dark energy,  (\ref{eq:sec:3.10-ede-coeffs}), one finds  that (\ref{eq:sec:eqn-becomes-ede-after-choice})   agrees with the equation of motion for $\xi^{i}$ given in   \cite{PhysRevD.76.023005}.  
\subsection{Scalar fields}
\label{sec:dsf}
As an explicit example, we can construct a theory where  $g_{\mu\nu}$ and $\nabla_{\mu}\phi$   enter the field content by combining into the kinetic scalar $\kin \defn - \half g^{\mu\nu} \nabla_{\mu}\phi\nabla_{\nu}\phi$, so that the field content is $\ld = \ld(\phi, \kin)$. In \tref{table:summaryparameters} we supplied   the coefficients for this general kinetic scalar field theory. The decoupling conditions (\ref{eq:sec:e0-decouple}, \ref{eq:sec:e0-decouple-4}) are trivially satisfied, the condition (\ref{eq:sec:e0-decouple-b}) becomes
\bea
-\bigg[ \dot{\rho} + 3 \hct(\rho +P)\bigg]=0,
\eea
which is always true, and (\ref{eq:sec:e0-decouple-c}) becomes
\bea
- \frac{2}{a}\bigg[ \ld_{,\kin}\ddot{\phi} + 2 \hct \ld_{,\kin}\dot{\phi} + \ld_{,\kin\kin}\dot{\phi}\dot{\kin} + \ld_{,\phi\kin}\dot{\phi}^2 - a^2\ld_{,\phi}\bigg]=0,
\eea
which is the Euler-Lagrange equation that one can compute directly and therefore vanishes identically. What this means is that for  all scalar field theories of the form $\ld = \ld(\phi, \kin)$, the $\xi^{\mu}$ and $\delta\phi$ fields decouple (i.e. one can set $\xi^{\mu}=0$ consistently). The components of the perturbed dark energy-momentum tensor become
\bse
\bea
\ep {U^0}_0 = \half \bigg( A_{\mathcal{V}} \delta\phi + \frac{1}{a}A_{\mathcal{Y}}\dot{\delta\phi}\bigg),\qquad
\ep {U^i}_0 = \frac{1}{2a} C_{\mathcal{Y}} \partial^i\delta\phi,
\eea
\bea
\ep {U^i}_j = \frac{1}{2a}\bigg( (\dot{C}_{\mathcal{Y}} + 3 \hct C_{\mathcal{Y}})\delta\phi + C_{\mathcal{Y}}\dot{\delta\phi}\bigg).
\eea
\ese
The $\nu=0$-component of the perturbed conservation law (\ref{eq:8.11-wow-a}) becomes
\bea
&&A_{\mathcal{Y}}\ddot{\delta\phi} + C_{\mathcal{Y}}\nabla^2\delta\phi +\bigg[  \dot{A}_{\mathcal{Y}} +aA_{\mathcal{V}}+(2A_{\mathcal{Y}}-3C_{\mathcal{Y}} )\hct \bigg]\dot{\delta\phi}\nonumber\\
&&\qquad + \bigg[ a\dot{A}_{\mathcal{V}} + 3a \hct A_{\mathcal{V}} - 3 \hct (\dot{C}_{\mathcal{Y}} + 3 \hct C_{\mathcal{Y}})\bigg]\delta\phi = a (\rho+P)\dot{h},
\eea
which is the evolution equation for the perturbed scalar field (we have verified that this reproduces known results) and only three functions are required to be specified: $A_{\mathcal{V}}, A_{\mathcal{Y}}, C_{\mathcal{Y}}$, in addition to the equation of state, $w$.

\section{Generalized perturbed fluid equations}
Here we study parameterizations of the generalized perturbed fluid equations. The main point of this section is to identify the entropy contribution, $\Gamma$, and anisotropic stresses, $\qsuprm{\Pi}{S}, \qsuprm{\Pi}{V}, \qsuprm{\Pi}{T}$, for the generalized dark theories we have discussed in this paper which will allow us to make connections with observations, and to deduce whether or not the form of $\Gamma$ suggested by   \cite{Weller:2003hw, PhysRevD.69.083503} is the most general way in which   entropy should be specified. We will continue to work in the synchronous gauge.

In   \cite{Ma:1995ey, PhysRevD.76.023005} the perturbed Bianchi identity, $\delta(\nabla_{\mu}U^{\mu\nu})=0$, is given for scalar perturbations in the synchronous gauge,
\bse
\label{eq:sec:2}
\bea
\dot{\delta} = - (1+w)\bigg( kv + \half \dot{h}\bigg) - 3 \hct w \Gamma,
\eea
\bea
\dot{v} = - \hct \big( 1 - 3w \big) v + \frac{w}{1+w}k\delta + \frac{k}{1+w}w\Gamma - \frac{2}{3}\frac{w}{1+w} k\qsuprm{\Pi}{S},
\eea
\ese
where $ w= P/\rho$ is the equation of state , $\delta  = \delta\rho/\rho$ is the density contrast, $\qsuprm{\Pi}{S}$ is the scalar anisotropic  source, and  the entropy contribution is given by
\bea
\label{eq:sec:gamma}
w\Gamma = \bigg( \frac{\delta P}{\delta\rho} - w \bigg) \delta,
\eea
where for simplicity we have assumed constant $w$. If the  entropy and anisotropic stress can be specified   in terms of the perturbed metric and fluid variables,
\bea
\label{eq:sec:eos-darksectorperts}
\Gamma = \Gamma(\delta, \dot{\delta}, v, \dot{v}, h,\eta),\qquad \qsuprm{\Pi}{S} = \qsuprm{\Pi}{S}(\delta, \dot{\delta}, v, \dot{v}, h,\eta),
\eea
then the   fluid equations (\ref{eq:sec:2})   become closed. The expressions (\ref{eq:sec:eos-darksectorperts}) constitute \textit{equations of state for dark sector perturbations}.

In \cite{Weller:2003hw, PhysRevD.69.083503}  the fluid equations (\ref{eq:sec:2}) are modified by introducing a gauge invariant density contrast. Their modification is equivalent to parameterizing the entropy as
\bea
\label{eq:Sec:lw-gamma-cssq}
w\Gamma = (\qsubrm{c}{s}^2 - w)\bigg( \delta + 3 \hct(1+w)\frac{v}{k}\bigg).
\eea
The important thing to notice here is that there is a single function which specifies the entropy: the sound speed $\qsubrm{c}{s}^2$. One should note that when $ w = -1$ the entropy contribution becomes $w\Gamma = (\qsubrm{c}{s}^2 +1) \delta$.

 It is standard practice to decompose the perturbed energy-momentum tensor with a density perturbation $\delta\rho$, velocity perturbation $v^i$, pressure perturbation $\delta P$ and anisotropic stresses $\pi_{ij}$,
\bea
\label{eq:sec1}
\delta {U^0}_0 = - \delta\rho,\qquad
\delta {U^i}_0 = - \rho(1+w)v^i,\qquad
\delta {U^i}_j = \delta P {\delta^i}_j + P{\pi^i}_j,
\eea
The tensor $\pi_{ij}$ is traceless, so that
\bea
\delta {U^i}_i =  3\delta P,\qquad \delta {U^i}_j - \frac{1}{3}{\delta^i}_j \delta {U^k}_k = P{\pi^i}_j.
\eea
This last expression is a useful way to determine the transverse-traceless components of a tensor.

The scalar  decomposition of the gauge field and metric perturbation is
\bse
\label{eq:sec:2.5-scaalr-decomp}
\bea
\xi_i = \partial_i \qsuprm{\xi}{S},\qquad h_{ij} = D_{ij}\eta + \frac{1}{3}h \delta_{ij},\qquad D_{ij} \defn \partial_i\partial_j - \frac{1}{3}\delta_{ij} \nabla^2,
\eea
where   $D_{ij}$ is a transverse-traceless spatial derivative operator, $ \nabla^2$ is the spatial Laplacian;  the trace of the metric perturbation is $h $ and the transverse-traceless part of the metric perturbation is
\bea
h_{ij} - \frac{1}{3}\delta_{ij} h = D_{ij}\eta.
\eea
The scalar decomposition of the perturbed fluid variables is
\bea
v^i = \partial^i\theta,\qquad \pi_{ij} = D_{ij} \qsuprm{\Pi}{S}.
\eea
\ese
We have that $\qsuprm{\Pi}{S}$ is the scalar anisotropic perturbation, and $\theta$ is the scalar velocity divergence perturbation. We will also decompose the vector and tensor pieces of vectorial and tensorial quantities in the usual way; the equations for the vector and tensor sectors are given in \cite{PhysRevD.76.023005}. To summarise our decompositions:
\bse
\bea
\xi^{\mu} \rightarrow \{\qsuprm{\xi}{S}, \qsuprm{\xi}{V}\},\qquad \ep g_{\mu\nu} \rightarrow \{h, \eta, \qsuprm{H}{V}, \qsuprm{H}{T}\},
\eea
\bea
 \ep {U^{\mu}}_{\nu} \rightarrow \{\delta, \theta, \delta P, \qsuprm{\Pi}{S}, \qsuprm{\theta}{V}, \qsuprm{\Pi}{V}, \qsuprm{\Pi}{T}\}.
\eea
\ese

As a final piece of notation, we will extract some of  the time dependance of a quantity by dividing out the density and write
\bea
\hat{X} \defn \frac{X}{\rho}.
\eea
A common example of this is using  an equation of state to link the pressure and density, $w =  {P}/{\rho}$. We will not \textit{a priori} assume that these ``hatted'' quantities are constant, but it may well turn out that the problem significantly simplifies if this is the case.
\subsection{No extra fields}
For the theory with field content $\ld = \ld(g_{\mu\nu})$,  one can use (\ref{eq:sec:no-eft-deu})  to   obtain the perturbed fluid variables
\bse
\label{eq:sec;2.8-1-2}
\bea
\delta  = -  \big(1 + \widehat{ \BW}{ } \big)\bigg( \nabla^2\qsuprm{\xi}{S}+ \half h\bigg),
\eea
\bea
\theta = \frac{1 -  \widehat{\CW}{ }}{  1+w}  {\qsuprm{\dot{\xi}}{S}},
\eea
\bea
\delta P =-\rho \bigg( \widehat{\DW}{ } + \frac{2}{3}\widehat{\EW}{ } + \frac{1}{3}w \bigg)\bigg( \nabla^2\qsuprm{\xi}{S} + \half h\bigg),
\eea
\bea
\qsuprm{\Pi}{S} = 2\frac{w-\widehat{\EW}}{w} \bigg( \qsuprm{\xi}{S} + \half \eta\bigg) ,
\eea
\bea
\qsuprm{\theta}{V} = \frac{1-\widehat{ \CW}}{1+w}\qsuprm{\dot{\xi}}{V},
\eea
\bea
\qsuprm{\Pi}{V} = \frac{\widehat{\EW}-w}{w}(k\qsuprm{\xi}{V}-\qsuprm{H}{V}),
\eea
\bea
\qsuprm{\Pi}{T} = \frac{w-\widehat{\EW}}{w}\qsuprm{H}{T}.
\eea
\ese
The entropy contribution (\ref{eq:sec:gamma}) can   be identified as
\bea
\label{eq:sec:2.13}
w\Gamma =\big( C - w\big)\delta,
\eea
where we defined
\bea
\label{eq:sec:defn-c-noef}
C \defn \frac{ {\DW}{ } + \frac{2}{3} {\EW}{ } + \frac{1}{3}P}{\rho + { \BW}{ }}.
\eea
Hence, we observe that this theory is capable of supporting non-trivial scalar, vector and tensor anisotropic and velocity sources.   When one makes the parameter choice corresponding to elastic dark energy (\ref{eq:sec:3.10-ede-coeffs}) one obtains $\Gamma =0$, which is in concordance with the results in  \cite{PhysRevD.76.023005}.

\subsection{Scalar fields}
For a theory with field content $\ld = \ld(g_{\mu\nu}, \phi, \nabla_{\mu}\phi)$, one can use (\ref{eq:sec:dsft-deu}) to obtain the perturbed fluid variables
\bse
\label{eq:sec:2.18-sft}
\bea
\delta  =-  \big(1 + \widehat{ \BW}{ } \big)\bigg(\nabla^2\qsuprm{\xi}{S}+ \half h\bigg)   -  \half \bigg( \widehat{A_{\mathcal{V}}} \delta\phi + \frac{1}{a}\widehat{A_{\mathcal{Y}} }\dot{\delta\phi}\bigg),
\eea
\bea
\theta = -\frac{1}{ 1+w}\bigg[(\widehat{\CW} - 1 ) \qsuprm{\dot{\xi}}{S}   + \frac{1}{2a} \widehat{C_{\mathcal{Y}}} \delta\phi\bigg],
\eea
\bea
\delta P =-\rho\bigg[  \widehat{ \DW} +\frac{2}{3}\widehat{\EW}+ \frac{1}{3}w \bigg]\bigg( \nabla^2\qsuprm{\xi}{S} + \half h\bigg)   -\frac{1}{2}\bigg( B_{\mathcal{V}}\delta\phi + \frac{1}{a}B_{\mathcal{Y}} \dot{\delta\phi}  \bigg) ,
\eea
\bea
\qsuprm{\Pi}{S} = 2\frac{w   - \widehat{\EW} }{w } \bigg( \qsuprm{\xi}{S} + \half \eta\bigg) ,
\eea
\bea
\qsuprm{\theta}{V} = \frac{1- \widehat{\CW}}{1+w}\qsuprm{\dot{\xi}}{V},
\eea
\bea
\qsuprm{\Pi}{V} = \frac{\widehat{\EW}-w}{w}(k\qsuprm{\xi}{V}-\qsuprm{H}{V}),
\eea
\bea
\qsuprm{\Pi}{T} =\frac{w   - \widehat{\EW} }{w } \qsuprm{H}{T}.
\eea
\ese
One should notice that this theory is capable of supporting anisotropic stresses, with $\qsuprm{\Pi}{S}, \qsuprm{\Pi}{V}, \qsuprm{\Pi}{T} \neq 0$ in general.  

The entropy contribution (\ref{eq:sec:gamma}) is found to be
\bea
\label{eq:sec:gamma-gen-1osft}
w\Gamma &=&  \bigg(    (1 + \widehat\BW) D-\mathcal{Z}  \bigg)\bigg( \nabla^2\qsuprm{\xi}{S} + \half h\bigg) -a(AD-B) (\widehat\CW -  1)  {\qsuprm{\dot{\xi}}{S}}{  }\nonumber\\
&&+ (D-w)\delta -a \big( AD-B   \big)      (1+w)\theta  ,  
\eea
where we have defined three frequently appearing ratios,
\bea
\label{eq:Sec:defn-dab}
D\defn \frac{B_{\mathcal{Y}}}{A_{\mathcal{Y}}} ,\qquad A\defn \frac{A_{\mathcal{V}}}{C_{\mathcal{Y}}},\qquad B\defn \frac{B_{\mathcal{V}}}{C_{\mathcal{Y}}},
\eea
and    the combination,
\bea
\mathcal{Z} \defn\frac{1}{3}\widehat P    +\widehat\DW + \frac{2}{3}\widehat\EW.
\eea

If we now apply the decoupling conditions (\ref{eq:sec:sft-decoup}) to (\ref{eq:sec:2.18-sft}, \ref{eq:sec:gamma-gen-1osft}, \ref{eq:Sec:defn-dab}), then we find
\bse
\label{eq:sec:sft-decoup-genfvars}
\bea
\delta = - \half \bigg( \widehat{A_{\mathcal{V}}} \delta\phi + \frac{1}{a}\widehat{A_{\mathcal{Y}} }\dot{\delta\phi}\bigg),
\eea
\bea
\theta =  -\frac{1}{ 2a(1+w)}  \widehat{C_{\mathcal{Y}}} \delta\phi,
\eea
\bea
\delta P= \frac{1}{2a}\bigg( (\dot{C}_{\mathcal{Y}} + 3 \hct C_{\mathcal{Y}})\delta\phi+C_{\mathcal{Y}} \dot{\delta\phi}  \bigg) ,
\eea
\bea
\label{eq:sec:fosft-eos-d}
\qsuprm{\Pi}{S} = \qsuprm{\theta}{V} = \qsuprm{\Pi}{V} = \qsuprm{\Pi}{T} = 0,
\eea
\bea
\label{eq:sec:fosft-eos-e}
w\Gamma  = (D-w)\delta - a (AD - B)(1+w)\theta,
\eea
\ese
where
\bea
D = - \frac{C_{\mathcal{Y}}}{A_{\mathcal{Y}}},\qquad A = \frac{A_{\mathcal{V}}}{C_{\mathcal{Y}}},\qquad B = - \frac{\dot{C}_{\mathcal{Y}} + 3\hct C_{\mathcal{Y}}}{aC_{\mathcal{Y}}}.
\eea
One should note that the anisotropic stress has now vanished. The perturbed fluid variables for theories with $\ld = \ld(\kin, \phi)$ are special cases of (\ref{eq:sec:sft-decoup-genfvars}).  The expressions (\ref{eq:sec:fosft-eos-d}, \ref{eq:sec:fosft-eos-e}) are examples of equations of state for dark sector perturbations.

\subsection{Example scalar field theories}
There are a number of explicit scalar field theories we will explicitly study which will enable us to get a feel for the typical   form of the functions which appear in the entropy. The theories we consider have Lagrangian densities given by
\bea
\ld = F(\kin)V(\phi),\qquad \ld = F(\kin) - V(\phi).
\eea
The first theory encompasses $k$-essence \cite{ArmendarizPicon:1999rj, ArmendarizPicon:2000ah}, and the second theory encompasses (for example) canonical scalar field theory.

It is useful to realize that the coefficients that appear in the decomposition of $\delta U^{\mu\nu}$ for a generic kinetic scalar field theory $\ld = \ld(\phi, \kin)$ can be written as
\bse
\bea
A_{\mathcal{V}} = -2 \big( 2\kin \ld_{,\kin\phi} - \ld_{,\phi}\big),\qquad B_{\mathcal{V}} =- 2\ld_{,\phi},
\eea
\bea
A_{\mathcal{Y}} = -2 \sqrt{2\kin}\big( 2\kin \ld_{,\kin\kin} + \ld_{,\kin}\big),\qquad B_{\mathcal{Y}} = - C_{\mathcal{Y}} = -2\ld_{,\kin}\sqrt{2\kin}.
\eea
and the energy density and pressure are computed via
\bea
\label{eq:Sec:rho-p-defn-ld-ldkin}
\rho = 2 \kin \ld_{,\kin} - \ld,\qquad P=\ld.
\eea
\ese
Our notation is $F' \defn \dd F/\dd \kin$, $V'\defn \dd V/\dd \phi$.

For a theory described by $\ld = F(\kin) - V(\phi)$ one obtains
\bse
\bea
w = - \bigg({1 -2\kin \frac{ F'}{F-V}}\bigg)^{-1},\qquad D = \bigg({1+2\kin \frac{F''}{F'}}\bigg)^{-1},
\eea
\bea
 AD -B = - \frac{2V'}{F'\sqrt{2\kin}}\bigg(1+2\kin\frac{  F''}{F'}\bigg)^{-1}\bigg(1 + \kin\frac{F''}{F'}\bigg) .
\eea
\ese
For the canonical scalar theory, $\ld = \kin - V(\phi)$, one obtains
\bea
D = 1,\qquad AD-B = - \frac{2V'}{\sqrt{2\kin}},
\eea
where   $ {V}'$ can be rewritten in terms of $\dot{\rho} = - 3\hct \rho(1+w)$ and $\dot{w}$, to obtain the formula
\bea
w\Gamma = (1-w)\bigg[ \delta - 3 \hct(1+w)\theta\bigg] - \dot{w}\theta.
\eea 
This is clearly of the same form as the expression suggested by  \cite{Weller:2003hw, PhysRevD.69.083503}, (\ref{eq:Sec:lw-gamma-cssq}) when we take $\qsubrm{c}{s}^2 = 1, \dot{w} = 0$.

For a theory described by $\ld = F(\kin)V(\phi)$ one obtains
\bse
\label{eq:sec:kessence-wdabd}
\bea
w = - \bigg({1 -2\kin \frac{ F'}{F}}\bigg)^{-1},\qquad D = \bigg({1+2\kin\frac{ F''}{F'}}\bigg)^{-1},
\eea
\bea
 AD-B = \frac{2V'}{F'\sqrt{2\kin}}\frac{F}{V}\bigg(1+2\kin\frac{  F''}{F'}\bigg)^{-1}\bigg[ {1+  \kin \bigg(\frac{F''}{F'}-\frac{F'}{F}\bigg)}{}\bigg].
\eea
\ese
It is interesting to note that   two of the frequently appearing combinations are
\bea
\frac{\kin F'}{F} = \frac{\dd\log F}{\dd\log \kin},\qquad \frac{\kin F''}{F'} = \frac{\dd\log F'}{\dd\log \kin},
\eea
which are the logarithmic slopes of $F$ and $F'$ respectively and bear a resemblance to ``slow-roll'' parameters.
For a pure $k$-essence theory, $\ld = F(\kin)$, we have that $V'=0$ and thus from (\ref{eq:sec:kessence-wdabd}) notice that  the entropy becomes $w\Gamma = (D - w)\delta$, which bears a strong resemblance to the form of the entropy in the no extra fields case (\ref{eq:sec:2.13}), but it is not of the form suggested by \cite{Weller:2003hw, PhysRevD.69.083503}.

It is possible to construct a Lagrangian density which has a specific equation of state. If we   impose $P = w\rho$ upon (\ref{eq:Sec:rho-p-defn-ld-ldkin}) then after trivial rearrangement we obtain 
\bea
\frac{1}{\ld} \pd{\ld}{\kin} = \frac{1+w}{2w\kin},
\eea
which can be integrated, yielding
\bea
\ld(\kin, \phi) = \tilde{V}(\phi)e^{\half \int \frac{1+w}{w}\frac{\dd\kin}{\kin}},
\eea
where $\tilde{V}(\phi)$ appears as a ``constant of integration''.
We can further impose $w_{,\kin}=0$, perform the integral, and find
\bea
\ld(\kin, \phi) = V(\phi)\kin^{\frac{1+w}{2w}}.
\eea
This theory has been constructed to have an equation of state $w$ which is independent of the kinetic term. In this case one is able to deduce that
\bea
D = w, \qquad AD-B = 0,
\eea
which means that the entropy for this particular theory vanishes.

\subsection{Generalized parameterization of entropy contributions}
Motivated by our results in the previous subsections, we   write down an expression for the parameterization of $w\Gamma$, which have been derived from the second order Lagrangian, and which we believe represents a wider range of theories than the parameterization of  \cite{Weller:2003hw, PhysRevD.69.083503}.  Our parameterization for the entropy contribution is
\bea
\label{eq:sec:gpoec-mainresult}
w\Gamma = (\alpha - w)\bigg[ \delta -  3\hct\beta(1+w)\theta\bigg],
\eea
where $ \alpha $ and $ \beta$ are both \textit{a priori} time-dependant  background quantities which we have explicitly determined for the no extra fields and  scalar field theories, although the hope would be that they would only be slowly varying functions of time.  We reiterate that (\ref{eq:sec:gpoec-mainresult}) constitutes an equation of state of dark sector perturbations. In \tref{table:entropy} we provide explicit formulae for these quantities for generic example theories. It should be noted that in the parameterization of \cite{Weller:2003hw, PhysRevD.69.083503}   $k$-essence is excluded, and the parameter $\alpha$ is always taken to be constant. It is rather simple to determine whether or not $\alpha$ can be expected to be constant or not for particular examples, as we show below.

Explicitly, for the theory $\ld = F(\kin) - V(\phi)$ one obtains
\bea
\label{eq:sec:alphabeta-theory1}
\alpha =  {}\bigg({1+ \frac{2\kin F''}{F'}}\bigg)^{-1},\qquad \beta = - \frac{1}{F'}\bigg[\frac{2aV'}{3\hct\sqrt{2\kin}}\bigg]\bigg(1 + \kin \frac{F''}{F}\bigg)\frac{\alpha}{\alpha -{w}{ }}.
\eea
We have written $\beta$ to isolate a   parameter combination, $  {2aV'}/({3\hct\sqrt{2\kin}} )$.

For the theory $\ld = F(\kin)V(\phi) $, one obtains
\bea
\alpha =  {}\bigg({1+ \frac{2\kin F''}{F'}}\bigg)^{-1},\qquad \beta =\frac{1}{F'} \bigg[\frac{2aV'}{3\hct \sqrt{2\kin}}\bigg]\frac{F}{V}\bigg({1+ \kin \bigg(\frac{F''}{F'} - \frac{F'}{F}\bigg)}{} \bigg)\frac{\alpha}{\alpha-  {w}{ }},\nonumber\\
\eea
which is the same as (\ref{eq:sec:alphabeta-theory1}) but with an extra factor in the expression for $\beta$.

\begin{table}[!h]
\begin{center}
\begin{tabular}{|c||c||c|}
\hline
Theory& $ \alpha $ & $ \beta$\\\hline
$\ld = \ld (g_{\mu\nu})$ &  $ C$ & 0\\
$\ld = \ld(\phi, \kin)$ & $D$ &$ \frac{AD- B}{3\hct(D- w  ) } a$ \\
$\ld = \kin - V(\phi)$ & 1 & $1$\\
$\ld = V(\phi)\kin^{\frac{1+w}{2w}}$ & $w$ & 0\\
$\ld = \ld(\kin)$ & $D$ & 0\\\hline
 \cite{Weller:2003hw, PhysRevD.69.083503} & $\qsubrm{c}{s}^2$ & $1$\\
\hline
\end{tabular}\caption{Collection of the quantities which determine the entropy contribution; the function $C$ is defined in (\ref{eq:sec:defn-c-noef}); in the final line we give the parameterization of  \cite{Weller:2003hw, PhysRevD.69.083503}.}\label{table:entropy}
\end{center}
\end{table}

\section{Modified gravity}
We will briefly discuss a class of theories which are more obviously ``modified gravity'' theories: we allow the dark sector to contain derivatives of the metric and no extra scalar fields. The simplest example is where the dark sectors field content is just the metric and its first derivative,
\bea
\ld = \ld(g_{\mu\nu}, \partial_{\alpha}g_{\mu\nu}).
\eea
The second order Lagrangian is given by
\bea
\label{eq:Sec:sec-olag-mmodgrav}
4\Diamond^2\ld =\half \mathcal{W}^{\mu\nu\alpha\beta} \lp g_{\mu\nu} \lp g_{\alpha\beta} + \mathcal{P}^{\mu\nu\rho\alpha\beta} \lp g_{\mu\nu}\nabla_{\rho}\lp g_{\alpha\beta} + \half\mathcal{Q}^{\sigma\mu\nu\rho\alpha\beta} \nabla_{\sigma} \lp g_{\mu\nu} \nabla_{\rho}\lp g_{\alpha\beta}.\nonumber\\
\eea
The tensors $\mathcal{W}, \mathcal{P}, \mathcal{Q}$ have the symmetries
\bea
\mathcal{W}^{\mu\nu\alpha\beta} = \mathcal{W}^{(\mu\nu)(\alpha\beta)}  = \mathcal{W}^{\alpha\beta\mu\nu} ,
\eea
\bea
\mathcal{P}^{\mu\nu\rho\alpha\beta} = \mathcal{P}^{(\mu\nu)\rho(\alpha\beta)},
\eea
\bea
\mathcal{Q}^{\sigma\mu\nu\rho\alpha\beta} = \mathcal{Q}^{\sigma(\mu\nu)\rho(\alpha\beta)} = \mathcal{Q}^{\rho\alpha\beta\sigma\mu\nu}.
\eea
This construction encompasses models which are often studied in the context of massive gravity theories (see e.g. the reviews \cite{Rubakov:2008nh, Hinterbichler:2011tt}). Explicitly computing the second measure-weighted variation of the Ricci scalar for metric perturbations about an arbitrary background yields
\bea
\label{eq:sec:diamond-2-ricci-gen}
\Diamond^2R&=&\nabla^{\lambda}\delta g_{\mu\nu}\bigg[  \half g^{\mu\nu} \nabla_{\lambda}{\delta g^{\alpha}}_{\alpha}+   \half \nabla_{\lambda}\delta g ^{\mu\nu}+   \nabla^{\mu}{\delta g^{\nu}}_{\lambda}-  g^{\mu\nu}\nabla^{\alpha}\delta g_{\alpha\lambda} \bigg]\nonumber\\
&& \qquad\qquad\qquad  +\frac{1}{4}\bigg[ 4R^{\alpha(\mu}g^{\nu)\beta} + 4R^{\beta(\mu}g^{\nu)\alpha}- 2 g^{\mu\nu}R^{\alpha\beta}-2 g^{\alpha\beta} R^{\mu\nu}  \nonumber\\
&&\qquad\qquad\qquad\qquad+ Rg^{\mu\nu}g^{\alpha\beta} - 2R g^{\mu(\alpha}g^{\beta)\nu}\bigg]\delta g_{\mu\nu}\delta g_{\alpha\beta}. 
\eea
This formula is clearly of the same form as (\ref{eq:Sec:sec-olag-mmodgrav}), but with $\mathcal{P}=0$,  and reproduces the relevant formulae in \cite{Battye:2004se} for   gravitational perturbations in a Minkowski background, and in \cite{Hinterbichler:2011tt} for perturbations in a Universe with constant curvature. 

Following the usual procedure, the perturbed dark energy-momentum tensor can be computed from (\ref{eq:Sec:sec-olag-mmodgrav}) and written as
\bea
\lp U^{\mu\nu} = \hat{\mathbb{W}}^{\mu\nu\alpha\beta}\lp g_{\alpha\beta},
\eea
where $\hat{\mathbb{W}}^{\mu\nu\alpha\beta}$ is a derivative operator which we write as
\bea
\hat{\mathbb{W}}^{\mu\nu\alpha\beta} = \mathbb{A}^{\mu\nu\alpha\beta} + \mathbb{B}^{\mu\nu\alpha\beta\rho}\nabla_{\rho} + \mathbb{C}^{\mu\nu\alpha\beta\rho\sigma}\nabla_{\rho}\nabla_{\sigma},
\eea
where one can identify
\bse
\bea
\mathbb{A}^{\mu\nu\alpha\beta} = - \half \bigg\{ \mathcal{W}^{\mu\nu\alpha\beta} + U^{\mu\nu}g^{\alpha\beta}- \nabla_{\rho}\mathcal{P}^{\alpha\beta\rho\mu\nu}\bigg\},
\eea
\bea
 \mathbb{B}^{\mu\nu\alpha\beta\rho}= - \half\bigg\{\mathcal{P}^{\mu\nu\rho\alpha\beta} - \mathcal{P}^{\alpha\beta\rho\mu\nu}   - \nabla_{\sigma}\mathcal{Q}^{\rho\alpha\beta\sigma\mu\nu}  \bigg\}
\eea
\bea
 \mathbb{C}^{\mu\nu\alpha\beta\rho\sigma} = \half \mathcal{Q}^{\sigma\alpha\beta\rho\mu\nu}.
\eea
\ese

A generic modified gravity theory containing curvature tensors is constructed from a field content  
\bea
\label{eq:sec:mod-grav-fieldcontent-1}
\ld = \ld(g_{\mu\nu}, R^{\alpha}_{\,\,\,\mu\beta\nu}),
\eea
where $R^{\alpha}_{\,\,\,\mu\beta\nu}$ is the Riemann tensor. The second order Lagrangian will have the schematic form
\bea
\Diamond^2\ld \sim \mathcal{W}\delta g \delta g + \mathcal{M}\delta g \delta R + \mathcal{N}\delta R \delta R,
\eea
where we have suppressed all indices for ease.
This field content will encompass theories containing arbitrary combinations of the Ricci tensor and scalar, and the Einstein tensor; popular examples of these types of theories include Gauss-Bonnet \cite{PhysRevD.67.024030, Clifton:2011jh}  and $F(R)$ gravities \cite{Capozziello:2003tk, PhysRevD.70.043528}.  To actually compute $\delta U^{\mu\nu}$ is  rather complicated and technical; we present some important steps in   Appendix \ref{section:append:fr} and the explicit form of $\delta U^{\mu\nu}$ for   $F(R)$ gravities. We find that the perturbed dark energy-momentum tensor can be  written as  
\bea
\delta U^{\mu\nu} =\hat{\mathbb{W}}^{\mu\nu\alpha\beta}\delta g_{\alpha\beta}
\eea
where  $\hat{\mathbb{W}}^{\mu\nu\alpha\beta}$ is a derivative operator which we write as
\bea
\hat{\mathbb{W}}^{\mu\nu\alpha\beta} =  \mathbb{A}^{\mu\nu\alpha\beta} +\mathbb{B}^{\mu\nu\alpha\beta\rho} \nabla_{\rho} + \ldots + \mathbb{E}^{\mu\nu\alpha\beta\rho\sigma\pi\zeta}\nabla_{\rho}\nabla_{\sigma}\nabla_{\pi}\nabla_{\zeta}.
\eea
 We will not present the explicit form of the coefficients $\mathbb{A}, \ldots, \mathbb{E}$ in generality in the decomposition of $\delta U^{\mu\nu}$ because they are   cumbersome and not particularly illuminating at this stage; however, in   Appendix \ref{section:append:fr} we provide the explicit coefficients for an $F(R)$ theory. The coefficients $\mathbb{A}, \ldots, \mathbb{E}$ can be written with an isotropic (3+1) decomposition to obtain all the parameters required.
 
We have only   sketched how our formalism  can be applied to theories containing high-order derivatives of the metric. However, it is clear that our formalism can be used to write down an effective Lagrangian density for the gravitational fluctuations which will automatically contain entire classes of known theories as well as encompassing theories which have never been considered before. As alluded to above, the second order Lagrangian (\ref{eq:sec:diamond-2-ricci-gen})   perhaps represents provide a fruitful theoretical testing ground for massive gravity theories.

\section{Discussion} 
In this paper we have outlined an approach to computing consistent perturbations to the generalized gravitational field equations for physically meaningful theories. The method requires a field content, but does not require a particular Lagrangian density to be presented for calculations to be performed. Once the field content has been specified we have shown how to write down an effective action for linearized perturbations. We imposed isotropy of the spatial sections so that all results we derived are compatible with an FRW metric.

We have provided an argument for why we use the Eulerian coordinate system to construct the perturbations which correspond to physically relevant quantities. The reason is that quantities in cosmology are perturbed about some known background, which is the same as the statement of using the Eulerian perturbation scheme.

We have given two   detailed examples illustrating how to use our formalism: one where the field content of the dark sector is entirely composed of the metric, and one where the dark sector contains a scalar field, its derivatives and the metric. \textit{A priori} we did not constrain these fields to combine into scalar quantities (such as a kinetic term).  We gave formulae for the effective Lagrangian for perturbations in these examples, which enabled us to write down the perturbed dark energy momentum tensor, $\lp {U^{\mu}}_{\nu}$. These expressions involved a number of coefficient-tensors which    can be   split with an isotropic (3+1) decomposition; the number of coefficients in this decomposition determines an upper bound on the number of functions which must be provided. The possible functions in the expansions   is further constrained by applying the perturbed conservation equation, $\delta (\nabla_{\mu}U^{\mu\nu})=0$; the number of possible functions can be further reduced by imposing, for example, a decoupling between the relevant field and the gauge field. In our appendices we have provided two explicit examples: kinetic scalar field theories with $\ld = \ld(\phi, \kin)$ and $F(R)$ gravities. Using these explicit calculations we have been able to justify our expansions of $\delta U^{\mu\nu}$. 

For the two examples we studied    we found that the number of parameters that were required to be specified depended upon what we imposed. (a) For the theory $\ld = \ld(g_{\mu\nu})$ at the level of the effective action there were 5 functions. Once we imposed the decoupling condition the 5 functions reduced to just 2. (b) For the theory $\ld = \ld(g_{\mu\nu},\phi, \nabla_{\mu}\phi)$ at the level of the effective action there were 14 functions. When we applied the linking conditions, the 14 functions reduced to 11.  When we imposed    reparameterization invariance by imposing the decoupling conditions, the 11 functions reduced to just 5,  although two of these are not present in the equations for cosmological perturbations in the synchronous gauge, and so we find that there are just 3 free functions. It is worth noting that all scalar field theories of the form $\ld = \ld(\phi, \kin)$ satisfy the decoupling conditions.

We also performed a study of the structure of the generalized perturbed fluid equations. For instance, we identified the entropy $w\Gamma$ and anisotropic stress sources $\Pi$ in our two general examples. We find that theories of the form $\ld = \ld(g_{\mu\nu})$ and $\ld = \ld(g_{\mu\nu},\phi, \nabla_{\mu}\phi)$ are both capable of supporting $w\Gamma \neq 0$ and $\Pi\neq 0$ when the theories are left in their general form.  When the decoupling conditions are applied to the theory with  $\ld(g_{\mu\nu},\phi, \nabla_{\mu}\phi)$  we find $\Pi=0$ ($\Gamma$ remains non-zero in general). Our main result is a general way to parameterize the entropy, (\ref{eq:sec:gpoec-mainresult}), which we derived from the second order Lagrangian.

We note that many of our results apply to theories with multiple scalar fields (see e.g. assisted inflation   \cite{Liddle:1998jc} or multi-field dark energy   \cite{Tsujikawa:2006mw, vandeBruck:2009gp, Frazer:2010zz}). For example, if a theory is has field content given by $\ld = \ld(g_{\mu\nu}, \supsm{\phi}{\rm A}, \nabla_{\mu}\supsm{\phi}{\rm A})$, then, regardless of the details of the theory (e.g.  even before imposing some generalized kinetic term, $\qsubrm{\kin}{gen} = - \half g^{\mu\nu}\qsubrm{G}{AB}\nabla_{\mu}\qsuprm{\phi}{A}\nabla_{\nu}\qsuprm{\phi}{B}$), the Lagrangian for perturbations will be given by
\bea
\sol &=& \subsm{\mathcal{A}}{\rm AB}\supsm{\vphi}{\rm{A}}\supsm{\vphi}{\rm{B}} +  \subsm{\mathcal{B}}{\rm AB}^{\mu}\supsm{\vphi}{\rm A} \nabla_{\mu}\supsm{\vphi}{\rm B}+\half  \subsm{\mathcal{C}}{\rm AB}^{\mu}\nabla_{\mu}\supsm{\vphi}{\rm A}\nabla_{\nu}\supsm{\vphi}{\rm B} \nonumber\\
&&+\frac{1}{4}\bigg[ \qsubrm{\mathcal{Y}}{A}^{\alpha\mu\nu} \nabla_{\alpha}\qsuprm{\vphi}{A} \lp g_{\mu\nu} + \qsubrm{\mathcal{V}}{A}^{\mu\nu}\qsuprm{\vphi}{A} \lp g_{\mu\nu} + \half \mathcal{W}^{\mu\nu\alpha\beta}\lp g_{\mu\nu}\lp g_{\alpha\beta}\bigg],
\eea
where repeated field-indices denote summation:
\bea
\qsubrm{\mathcal{V}}{A}^{\mu\nu}\qsuprm{\vphi}{A}  =\sum^{\scriptscriptstyle\rm{N}}_{{\scriptscriptstyle\rm A}=1}\qsubrm{\mathcal{V}}{A}^{\mu\nu}\qsuprm{\vphi}{A} .
\eea
The generalized perturbed fluid variables can then be worked out, where one would find that the expressions we provided in (\ref{eq:sec:2.18-sft}) still hold, except that we would need to identify
\bea
A_{\mathcal{V}}\vphi = A_{\mathcal{V}\scriptscriptstyle\rm{A}}\qsuprm{\vphi}{A},\qquad B_{\mathcal{V}}\vphi = B_{\mathcal{V}\scriptscriptstyle\rm{A}}\qsuprm{\vphi}{A},
\eea
\bea
 A_{\mathcal{Y}}\dot{\vphi} = A_{\mathcal{Y}\scriptscriptstyle\rm{A}}\qsuprm{\dot{\vphi}}{A},\qquad B_{\mathcal{Y}}\dot{\vphi} = B_{\mathcal{Y}\scriptscriptstyle\rm{A}}\qsuprm{\dot{\vphi}}{A}\qquad C_{\mathcal{Y}} {\vphi} = C_{\mathcal{Y}\scriptscriptstyle\rm{B}}\qsuprm{ {\vphi}}{A}.
\eea
What this means is that, for example, the parameterization of the entropy     (\ref{eq:sec:gpoec-mainresult}) also   applies to these multi-field models.

 We will sketch a formulation involving vector fields here, but a full study will be presented in a follow up paper. For a field content containing a vector field $A^{\mu}$ and its derivatives,
\bea
\ld = \ld(A^{\mu}, \nabla_{\alpha}A^{\mu},\nabla_{\alpha}\nabla_{\beta}A^{\mu},\ldots),
\eea
the Lagrangian perturbed dark energy-momentum tensor would take on the form
\bea
\lp U^{\mu\nu} = \hat{\mathbb{Z}}^{\mu\nu}_{\quad \alpha}\lp A^{\alpha},
\eea
where $\hat{\mathbb{Z}}^{\mu\nu}_{\quad \alpha}$ is an operator which we would expand as
\bea
\hat{\mathbb{Z}}^{\mu\nu}_{\quad \alpha} =  {\mathbb{P}}^{\mu\nu}_{\quad \alpha} +  {\mathbb{Q}}^{\mu\nu\quad \beta}_{\quad \alpha}\nabla_{\beta} + {\mathbb{R}}^{\mu\nu\quad \beta\gamma}_{\quad \alpha}\nabla_{\beta}\nabla_{\gamma} + \ldots
\eea
%
%
%

The formalism we have developed can be used as a stepping-stone to create a tool which can be used to discriminate different gravity and dark energy theories using  experiment. All perturbations in our formalism have a obvious, consistent and concrete origin from an effective action.
 
\section*{Acknowledgements}
 We have benefitted from  conversations with Tessa Baker, Pedro Ferreira, Anthony Lewis, Jochen Weller, Adam Moss, Rachel Bean and  Alkistis Pourtisdou.  
\appendix
  
\section{Kinetic scalar fields}
\label{appendic:darkfluids}
Here we will give the explicit forms of the tensors that appear in the dark scalar field case where only first order   derivatives appear in the Lagrangian density of the dark sector. We will study the explicit   case   where the metric $g_{\mu\nu}$ and the derivative of the scalar field $\nabla_{\mu }\phi$ only appear in the dark sector through the kinetic term,
\bea
\kin = -\half g^{\mu\nu} \nabla_{\mu}\phi \nphid{\nu}.
\eea
Hence, the field content of a first order scalar field theory with a kinetic term can be written as
\bea
\ld = \ld(\phi, \kin),
\eea
the generalized gravitational action is
\bea
S = \int \dd^4x\sqrt{-g} \, \bigg[ R + 16\pi G \qsubrm{\ld}{m} +\ld(\phi, \kin) \bigg],
\eea
and the generalized gravitational field equations are given by $G^{\mu\nu} = 8 \pi G T^{\mu\nu} + U^{\mu\nu}$ where the dark energy-momentum tensor is given by
\bea
\label{eq:sec:deltaumunu-scalarfld-umunu}
U^{\mu\nu}= \ld_{, \mathcal{X}} \nabla^{\mu}\phi\nabla^{\nu}\phi + \ld g^{\mu\nu}.
\eea 


By explicit calculation we can identify the   quantities $\{\mathcal{A}, \mathcal{B}^{\mu},\mathcal{C}^{\mu\nu}, \mathcal{V}^{\mu\nu},\mathcal{Y}^{\alpha\mu\nu}, \mathcal{W}^{\mu\nu\alpha\beta}\}$ which we introduced in our second order Lagrangian (\ref{eq:sec:4.1-Diald-dakf}). To aid our calculation  it is useful to realize that the variations of the Lagrangian are
\bse
\bea
\delta\ld = \ld_{,\phi}\delta\phi + \ld_{,\kin}\delta\kin,
\eea
\bea
\delta^2\ld = \ld_{,\phi\phi}(\delta\phi)^2 + \ld_{,\kin\kin}(\delta\kin)^2 + 2 \ld_{,\phi\kin}\delta\phi\delta\kin + \ld_{,\kin}\delta^2\kin,
\eea
and those of the  kinetic term are
\bea
\delta\kin = \half \delta g_{\mu\nu}\nabla^{\mu}\phi\nabla^{\nu}\phi- \nabla^{\mu}\phi \nabla_{\mu}\delta\phi,
\eea
\bea
\delta^2\kin &=& - g^{\mu\nu}\nabla_{\mu}\delta\phi\nabla_{\nu}\delta\phi + 2 g^{\alpha(\mu}\nabla^{\nu)}\phi\nabla_{\alpha}\delta\phi \delta g_{\mu\nu} \nonumber\\
&&- \half \bigg[ g^{\mu(\alpha}\nabla^{\beta)}\phi\nabla^{\nu}\phi + g^{\nu(\alpha}\nabla^{\beta)}\phi\nabla^{\mu}\phi\bigg] \delta g_{\mu\nu}\delta g_{\alpha\beta}.
\eea
\ese
We now combine these expressions to form $\Diamond^2\ld$, as defined in (\ref{eq:2.14b-d2ld}) and compare the result with (\ref{eq:sec:4.1-Diald-dakf}). By appropriate identifications, one   finds that
\bea
\label{eq:sec:append-sol-1osft-abc}
\mathcal{A} = - \half \ld_{,\phi\phi},\qquad \mathcal{B}^{\mu} = \ld_{,\phi\kin}\nabla^{\mu}\phi,\qquad \mathcal{C}^{\mu\nu} = \ld_{,\kin}g^{\mu\nu} - \ld_{,\kin\kin} \nabla^{\mu}\phi\nabla^{\nu}\phi,
\eea
\bse
\label{eq:sec:deltaumunu-scalarfld}
\bea
 \mathcal{V}^{\mu\nu} =  -2 \bigg[ g^{\mu\nu} \ld_{,\phi} + \ld_{,\kin\phi}\nphiu{\mu}\nphiu{\nu}\bigg],
\eea
\bea
\mathcal{Y}^{\alpha\mu\nu} &=&    2 \bigg[  \ld_{,\kin\kin}\nphiu{\alpha}\nphiu{\mu}\nphiu{\nu}+ \ld_{,\kin}\bigg(g^{\mu\nu}\nphiu{\alpha}- 2g^{\alpha(\mu}\nphiu{\nu)}\bigg)   \bigg],
\eea
\bea
\mathcal{W}^{\alpha\beta\mu\nu} &=& -\ld_{,\kin\kin}\nphiu{\mu}\nphiu{\nu}\nphiu{\alpha}\nphiu{\beta} - \ld_{,\kin} \bigg( g^{\mu\nu}\nphiu{\alpha}\nphiu{\beta}+ g^{\alpha\beta}\nphiu{\mu}\nphiu{\nu}\bigg) \nonumber\\
&&+2 \ld_{,\kin} \bigg( g^{\mu(\alpha}\nphiu{\beta)}\nphiu{\nu}+ g^{\nu(\alpha}\nphiu{\beta)}\nphiu{\mu}\bigg) - \ld\bigg( g^{\mu\nu}g^{\alpha\beta} - 2 g^{\mu(\alpha}g^{\beta)\nu}\bigg).\nonumber\\
\eea
\ese
The perturbed dark energy-momentum tensor can be written as
\bea
\lp U^{\mu\nu} &=&  -\half \bigg\{   \mathcal{V}^{\mu\nu}\delta\phi +  \mathcal{Y}^{\alpha\mu\nu} \nabla_{\alpha}\delta\phi \bigg\}    -\half \bigg\{\mathcal{W}^{\alpha\beta\mu\nu}  +  g^{\alpha\beta} {} U^{\mu\nu} \bigg\}\lp g_{\alpha\beta}.
\eea
One can use (\ref{eq:sec:append-sol-1osft-abc}) to compute, for example, the Euler-Lagrange equation (\ref{eq:el-4.2-}) which was computed from the second order Lagrangian. For a canonical theory it is simple to use (\ref{eq:sec:append-sol-1osft-abc}) to obtain the well known formula $\square\delta\phi + V''\delta\phi = \ep S$. This vindicates our use of the second order Lagrangian as the Lagrangian for the perturbed scalar field.

 We now decompose the tensors (\ref{eq:sec:deltaumunu-scalarfld}) with an isotropic $(3+1)$-split; we  write
\bea
\nphid{\mu}= -\dot{\phi} u_{\mu},\qquad g_{\mu\nu} = \gamma_{\mu\nu} - u_{\mu}u_{\nu},\qquad u^{\mu}u_{\mu} = -1,\qquad u^{\mu}\gamma_{\mu\nu}=0.
\eea
 Notice that with our signature choice, the definition of the covariant derivative of the scalar field   means that $\nabla_{\mu}\phi = - \dot{\phi} u_{\mu}\rightarrow \partial_t\phi = \dot{\phi}$. The choice of using the minus sign is only important for terms linear or cubic in derivatives of $\phi$; i.e. for $\mathcal{Y}^{\alpha\mu\nu}$. The energy-momentum tensor (\ref{eq:sec:deltaumunu-scalarfld-umunu}) can now be written as
\bea
U^{\mu\nu} =  {\rho}{ } u^{\mu}u^{\nu}+ {P}{ }  \gamma^{\mu\nu} ,
\eea
where   the energy density and pressure are given by
\bea
\label{eq:sec4.18-en-press-decomp}
 {\rho}{ } = \ld_{,\kin}\dot{\phi}^2- \ld ,\qquad  {P}{ } =  \ld .
\eea
Inserting the   (3+1) split into the perturbed energy momentum tensor (\ref{eq:sec:deltaumunu-scalarfld}) and comparing with (\ref{eq:sec-4.5-decomp-vy-df}), one finds that 
\bse
\bea
A_{\mathcal{V}} = -2\big(\ld_{,\kin\phi} \dot{\phi}^2 - \ld_{,\phi}\big),\qquad
B_{\mathcal{V}} = -2 \ld_{,\phi} ,
\eea
\bea
A_{\mathcal{Y}} =  -2 \bigg[ \ld_{,\kin\kin} \dot{\phi}^3 + \ld_{,\kin}\dot{\phi}\bigg],\qquad B_{\mathcal{Y}} = -C_{\mathcal{Y}}= -2 \ld_{,\kin}\dot{\phi} ,
\eea
\bea
 {A}_{\mathcal{W}} =  -\bigg[\ld_{,\kin\kin}\dot{\phi}^4 + 2 {\rho}{ } + {P}{ }\bigg],\qquad
 {B}_{\mathcal{W}}  = - {C}_{\mathcal{W}} =-  {\rho}{ },\qquad \DW = - \EW =-  {P}{ }\nonumber\\
\eea
\ese
The sets of  coefficients $(A_{\mathcal{V}}, B_{\mathcal{V}}),(A_{\mathcal{Y}}, B_{\mathcal{Y}})$ are in general different, however they only are different in  non-canonical theories where there is an explicit coupling between the scalar field and its kinetic term and where a non-trivial function of the kinetic term appears in the Lagrangian density. It is also interesting to note that coefficients $\DW $ and $\EW $ which appear in the general decomposition of $\mathcal{E}_{\mu\nu\alpha\beta}$ (\ref{eq:e-gen-metonly-3.4c}) are found to be equal but opposite in this dark fluid scenario (in fact, their values are set to the pressure). Similarly, $B_{\mathcal{Y}} = - C_{\mathcal{Y}}$.

 
\section{$F(R)$ and Gauss-Bonnet gravities}
\label{section:append:fr}
We will   compute the field equations for a generalized modified gravity theory, and inparticular to identifying the dark energy-momentum tensor $U^{\mu\nu}$. The class of theories we will consider is defined by the action
\bea
S = \int \dd^4x\, \sqrt{-g} \bigg[ R +2 \qsubrm{\ld}{mg}(g_{\mu\nu}, R^{\alpha}_{\,\,\,\mu\beta\nu}) - 16 \pi G \qsubrm{\ld}{m}\bigg].
\eea
We will now calculate $U^{\mu\nu}$ for this theory. To do so, it is useful to define the derivatives of the Lagrangian to be
\bea
A^{\mu\nu} \defn \frac{\delta\qsubrm{\ld}{mg}}{\delta g_{\mu\nu}},\qquad B_{\mu}^{\,\,\,\alpha\nu\beta} \defn \frac{\delta \qsubrm{\ld}{mg}}{\delta R^{\mu}_{\,\,\,\alpha\nu\beta}},
\eea
and note that the variations of the Riemann and Ricci tensor and scalar can be written as
\bse
\bea
\label{eq:sec:identity-deltaR}
\delta R_{\mu\nu} = g^{\beta\sigma}g_{\sigma\alpha}\delta R^{\alpha}_{\,\,\,\mu\beta\nu},\qquad \delta R = g^{\alpha\beta}g^{\rho\sigma}g_{\sigma\pi}\delta R^{\pi}_{\,\,\,\alpha\rho\beta} - R^{\alpha\beta}\delta g_{\alpha\beta},
\eea
\bea
\delta R^{\alpha}_{\,\,\,\mu\beta\nu} = \Theta^{\alpha\xi\rho\sigma\pi}_{\qquad\,\,\,\mu\beta\nu} \nabla_{\xi}\nabla_{\rho}\delta g_{\sigma\pi},
\eea
\ese
where, for convenience we have defined
\bea
\label{eq:sec:defn-Theta-genmodgrav}
\Theta^{\alpha\xi\rho\sigma\pi}_{\qquad\,\,\,\mu\beta\nu} \defn g^{\alpha\rho}\delta^{\sigma}_{\mu}\delta^{[\pi}_{\beta}\delta^{\xi]}_{\nu} + g^{\alpha\pi}\delta^{\sigma}_{\mu}\delta^{[\rho}_{\nu} \delta^{\xi]}_{\beta} + g^{\alpha\pi} \delta^{\rho}_{\mu}\delta^{[\sigma}_{\nu}\delta^{\xi]}_{\beta}.
\eea
Thus, varying the Lagrangian yields
\bea
\delta \qsubrm{\ld}{mg} = \bigg(A^{\sigma\pi} + \Theta^{\alpha\xi\rho\sigma\pi}_{\qquad\,\,\,\mu\beta\nu} \nabla_{\rho}\nabla_{\xi}B_{\alpha}^{\,\,\,\mu\beta\nu}\bigg)\delta g_{\sigma\pi} + \nabla_{\mu}\delta S^{\mu},
\eea
where the term which will only contribute to a surface integral is given by
\bea
\delta S^{\rho} \defn B_{\alpha}^{\,\,\,\mu\beta\nu} \Theta^{\alpha\rho\xi\sigma\pi}_{\qquad\,\,\,\mu\beta\nu}\nabla_{\xi} \delta g_{\sigma\pi} - \Theta^{\alpha\xi\rho\sigma\pi}_{\qquad\,\,\,\mu\beta\nu} \nabla_{\xi}(\delta g_{\sigma\pi}B_{\alpha}^{\,\,\,\mu\beta\nu}).
\eea

The field equations are given by $G^{\mu\nu}  = 8 \pi GT^{\mu\nu} + U^{\mu\nu}$, where, under the usual definition, the dark energy momentum tensor is given by
\bea
\label{eq:sec:u-gen-modgrav}
U^{\sigma\pi} = - \bigg[ g^{\sigma\pi} \qsubrm{\ld}{mg} +2A^{\sigma\pi} + 2\Theta^{\alpha\xi\rho\sigma\pi}_{\qquad\,\,\,\mu\beta\nu} \nabla_{\rho}\nabla_{\xi}B_{\alpha}^{\,\,\,\mu\beta\nu}\bigg].
\eea

For an $F(R)$ theory we can use (\ref{eq:sec:identity-deltaR}) to obtain
\bea
\label{eq:sec:fr-ab}
A^{\mu\nu} = -\ld_{,R}R^{\mu\nu},\qquad B_{\pi}^{\,\,\,\alpha\rho\beta} = \ld_{,R}g^{\alpha\beta}\delta^{\rho}_{\pi}.
\eea
For a consistency check, an $F(R)$ theory has $\qsubrm{\ld}{mg} = \half f(R)$. Using (\ref{eq:sec:fr-ab}) to calculate the final term in (\ref{eq:sec:u-gen-modgrav})  yields
\bea
\Theta^{\alpha\xi\rho\sigma\pi}_{\qquad\,\,\,\mu\beta\nu}\delta^{\beta}_{\alpha}g^{\mu\nu} = g^{\rho\sigma}g^{\xi\pi} - g^{\rho\xi}g^{\pi\sigma},
\eea
so that we obtain
\bea
U^{\sigma\pi} = (R^{\sigma\pi} + g^{\sigma\pi}\square - \nabla^{\sigma}\nabla^{\pi})f_{,R} - \half g^{\sigma\pi}f,
\eea
which is identical to the expression one finds through direct calculation. 

For a (generalized) Gauss-Bonnet theory, the action is given in terms of the Gauss-Bonnet term, $\mathcal{G}$, 
\bea
S = \int \dd^4x\, \sqrt{-g} \ld(\mathcal{G}),\qquad \mathcal{G} \defn R^2 - 4 R^{\mu\nu}R_{\mu\nu} + 4R^{\mu\nu\alpha\beta}R_{\mu\nu\alpha\beta}.
\eea
It is useful to realize   that the variation of the Gauss-Bonnet term  $\delta\mathcal{G}$ can be written as
\bea
\delta \mathcal{G} = \mathfrak{A}^{\epsilon\kappa}\delta g_{\epsilon\kappa}+ \mathfrak{B}_{\lambda}^{\,\,\,\gamma\epsilon\kappa} \delta R^{\lambda}_{\,\,\,\gamma\epsilon\kappa},
\eea
where
\bse
\bea
\mathfrak{A}^{\epsilon\kappa} &=& - 2 RR^{\epsilon\kappa} + 8R^{\alpha\epsilon}{R^{\kappa}}_{\alpha}+R^{\epsilon\nu\alpha\beta}R^{\kappa}_{\,\,\,\nu\alpha\beta} - R_{\xi}^{\,\,\,\nu\alpha\kappa}R^{\xi\quad\epsilon}_{\,\,\,\nu\alpha} \nonumber\\
&&-R_{\xi}^{\,\,\,\nu\kappa\alpha}R^{\xi\,\,\,\epsilon}_{\,\,\,\nu\,\,\,\alpha}-R_{\xi}^{\,\,\,\kappa\alpha\beta}R^{\xi\epsilon}_{\,\,\,\,\,\,\alpha\beta}, 
\eea
\bea
\mathfrak{B}_{\lambda}^{\,\,\,\gamma\epsilon\kappa} &=& 2\bigg(Rg^{\gamma\kappa}  - 4 R^{\gamma\kappa}\bigg){\delta^{\epsilon}}_{\lambda}     +2R_{\lambda}^{\,\,\,\gamma\epsilon\kappa}.
\eea
\ese
Hence, for a Gauss-Bonnet theory, we find that
\bea
A^{\mu\nu} = \ld_{,\mathcal{G}} \mathfrak{A}^{\mu\nu},\qquad B_{\mu}^{\,\,\,\alpha\nu\beta} = \ld_{,\mathcal{G}} \mathfrak{B}_{\mu}^{\,\,\,\alpha\nu\beta} .
\eea
To calculate   $\delta U^{\mu\nu}$ in generality is   complicated and cumbersome to write down.

 In general it is   possible to write $\delta U^{\mu\nu}$ as a single   rank-4 pseudo-tensor operator acting on $\delta g^{\alpha\beta}$, so that
\bse
\label{eq:Sec:deltau_append_modgrav_proto}
\bea
\delta U^{\mu\nu} = \hat{\mathbb{W}}^{\mu\nu\alpha\beta} \delta g_{\alpha\beta},
\eea 
where $ \hat{\mathbb{W}}^{\mu\nu\alpha\beta} $ is an operator given by
\bea
 \hat{\mathbb{W}}^{\mu\nu\rho\sigma} &=& \mathbb{A}^{\mu\nu\rho\sigma}  + \mathbb{B}^{\mu\nu\rho\sigma\alpha}\nabla_{\alpha}  + \mathbb{C}^{\mu\nu\rho\sigma\alpha\beta} \nabla_{\alpha}\nabla_{\beta}  \nonumber\\
&&\qquad \qquad + \mathbb{D}^{\mu\nu\rho\sigma\lambda\alpha\beta} \nabla_{\lambda}\nabla_{\alpha}\nabla_{\beta}  +\mathbb{E}^{\mu\nu\rho\sigma\zeta\lambda\alpha\beta} \nabla_{\zeta}\nabla_{\lambda}\nabla_{\alpha}\nabla_{\beta} .
\eea
\ese
The coefficients $\mathbb{A}\cdots \mathbb{E}$ can be written using the (3+1) split.

We will   explicitly calculate $\delta U^{\mu\nu}$ for $F(R)$ gravities, and show that $\delta U^{\mu\nu}$ is indeed of this form. The action we study is given by
\bea
S = \int \dd^4x\, \sqrt{-g} \bigg[ R + f(R) - 16\pi G \qsubrm{\ld}{m}\bigg].
\eea
The dark energy-momentum tensor  is given by
\bea
U^{\mu\nu}=\bigg(R^{\mu\nu} + g^{\mu\nu}\square - \nabla^{\mu}\nabla^{\nu} \bigg)f'- \half fg^{\mu\nu} .
\eea
Perturbing this yields
\bea
\label{eq:sec:deltau_fr}
\delta U^{\mu\nu} = \mathfrak{A}^{\mu\nu\rho\sigma}\delta g_{\rho\sigma} + \mathfrak{B}^{\mu\nu} \delta R + \mathfrak{C}^{\alpha\mu\nu}\nabla_{\alpha}\delta R + \mathfrak{D}^{\alpha\beta\rho\sigma}\nabla_{\alpha}\nabla_{\beta} \delta R + {\mathfrak{E}_{\alpha}}^{\rho\sigma\mu\nu}\delta \cs{\alpha}{\rho}{\sigma},\nonumber\\
\eea 
where we have defined
\bse
\bea
\mathfrak{A}^{\mu\nu\rho\sigma} &\defn& - 2 f'g^{\rho(\mu}R^{\nu)\sigma} - g^{\mu\rho}g^{\nu\sigma}\square f' - f'''g^{\mu\nu}\nabla^{\rho}R\nabla^{\sigma}R + 2f'''g^{\rho(\mu}\nabla^{\nu)}R\nabla^{\sigma}R  \nonumber\\
&&+\half f g^{\mu\rho}g^{\nu\sigma} - f''g^{\mu\nu}\nabla^{\rho}\nabla^{\sigma}R + 2 f'' g^{\rho(\mu}\nabla^{\nu)}\nabla^{\sigma}R,
\eea
\bea
\mathfrak{B}^{\mu\nu} \defn f'' g^{\mu\nu} + f''' g^{\mu\nu}\square R + f'''' g^{\mu\nu}\nabla^{\alpha}R\nabla_{\alpha}R - f'''' \nabla^{\mu}R\nabla^{\nu}R - \half f' g^{\mu\nu},
\eea
\bea
\mathfrak{C}^{\alpha\mu\nu} \defn 2f''' \bigg( g^{\mu\nu}\nabla^{\alpha}R - g^{\alpha(\mu}\nabla^{\nu)}R\bigg),
\eea
\bea
\mathfrak{D}^{\alpha\beta\mu\nu} \defn f'' \bigg(g^{\mu\nu}g^{\alpha\beta} - g^{\mu\alpha}g^{\nu\beta}\bigg),
\eea
\bea
{\mathfrak{E}_{\alpha}}^{\rho\sigma\mu\nu} \defn f'' \bigg( g^{\mu\rho}g^{\nu\sigma} - g^{\mu\nu}g^{\rho\sigma}\bigg) \nabla_{\alpha}R
\eea
\ese

To actually compute  the explicit form of $\hat{\mathbb{W}}^{\mu\nu\alpha\beta} $ is a rather convoluted task and to do so we use the fact that perturbations such as $\delta R, \delta R_{\mu\nu}, \delta \cs{\lambda}{\alpha}{\beta}$ can be written entirely in terms of $\delta g^{\alpha\beta}$.  For example, we have the identities
\bea
\delta \cs{\rho}{\mu}{\nu} = \half g^{\rho\sigma}\bigg( \nabla_{\mu}\delta g_{\nu\sigma} + \nabla_{\nu}\delta g_{\mu\sigma} - \nabla_{\sigma}\delta g_{\mu\nu}\bigg) ,
\eea
\bea
\delta R_{\alpha\beta} = \nabla_{\rho}\delta \cs{\rho}{\alpha}{\beta} - \nabla_{\beta} \delta \cs{\rho}{\alpha}{\rho}  ,
\eea
which can be rewritten into the more useful form
\bse
\label{eq:sec:append_useful_decs-dr}
\bea
\delta \cs{\rho}{\mu}{\nu} =  \bigg( g^{\rho\pi}\delta^{\xi}_{(\mu}\delta^{\lambda}_{\nu)} - \half g^{\rho\xi}\delta^{\lambda}_{\mu}\delta^{\pi}_{\nu}\bigg) \nabla_{\xi}\delta g_{\lambda\pi}, 
\eea
\bea
\delta R_{\alpha\beta}  ={\Theta^{\gamma\lambda\xi\epsilon\pi}}_{\alpha\gamma\beta}\nabla_{\lambda}\nabla_{\xi}\delta g_{\epsilon\pi},
\eea
\ese
where ${\Theta^{\gamma\lambda\xi\epsilon\pi}}_{\alpha\gamma\beta}$ is defined in (\ref{eq:sec:defn-Theta-genmodgrav}).
By writing $\delta R = g^{\mu\nu} \delta R_{\mu\nu} - R^{\mu\nu} \delta g_{\mu\nu}$  we find that from (\ref{eq:sec:deltau_fr}) we can obtain
\bea
\delta U^{\mu\nu} &=& \bigg[ \mathfrak{A}^{\mu\nu\rho\sigma} - \mathfrak{B}^{\mu\nu}R^{\rho\sigma} - \mathfrak{C}^{\alpha\mu\nu}\nabla_{\alpha}R^{\rho\sigma} - \mathfrak{D}^{\alpha\beta\mu\nu} \nabla_{\alpha}\nabla_{\beta} R^{\rho\sigma} \bigg]\delta g_{\rho\sigma}+\bigg[ {\mathfrak{E}_{\alpha}}^{\rho\sigma\mu\nu}\bigg] \delta \cs{\alpha}{\rho}{\sigma}\nonumber\\
&&+ \bigg[ - \mathfrak{C}^{\alpha\mu\nu}R^{\rho\sigma} - 2\mathfrak{D}^{(\alpha\beta)\mu\nu}\nabla_{\beta}R^{\rho\sigma} \bigg]\nabla_{\alpha}\delta g_{\rho\sigma} + \bigg[-\mathfrak{D}^{\alpha\beta\mu\nu}R^{\rho\sigma}\bigg] \nabla_{\alpha}\nabla_{\beta}\delta g_{\rho\sigma} \nonumber\\
&&+ \bigg[ \mathfrak{B}^{\mu\nu} g^{\alpha\beta}\bigg] \delta R_{\alpha\beta} + \bigg[ \mathfrak{C}^{\zeta\mu\nu}g^{\alpha\beta} \bigg]\nabla_{\zeta}\delta R_{\alpha\beta}  + \bigg[ \mathfrak{D}^{\zeta\kappa\mu\nu} g^{\alpha\beta}\bigg]\nabla_{\zeta}\nabla_{\kappa} \delta R_{\alpha\beta},
\eea 
and if we insert (\ref{eq:sec:append_useful_decs-dr}) into this we obtain
\bea
\delta U^{\mu\nu} &=& \bigg\{ \mathfrak{A}^{\mu\nu\rho\sigma} - \mathfrak{B}^{\mu\nu}R^{\rho\sigma} - \mathfrak{C}^{\alpha\mu\nu}\nabla_{\alpha}R^{\rho\sigma} - \mathfrak{D}^{\alpha\beta\mu\nu} \nabla_{\alpha}\nabla_{\beta} R^{\rho\sigma} \bigg\}\delta g_{\rho\sigma}\nonumber\\
&&+\bigg\{ {\mathfrak{E}_{\xi}}^{\lambda\beta\mu\nu} \big[ g^{\xi\sigma}\delta^{\alpha}_{(\lambda}\delta^{\rho}_{\pi)} - \half g^{\xi\alpha}\delta^{\rho}_{\lambda} \delta^{\sigma}_{\pi} \big]   - \mathfrak{C}^{\alpha\mu\nu}R^{\rho\sigma} - 2\mathfrak{D}^{(\alpha\beta)\mu\nu}\nabla_{\beta}R^{\rho\sigma}  \bigg\}\nabla_{\alpha}\delta g_{\rho\sigma}  \nonumber\\
&&+\bigg\{  \mathfrak{B}^{\mu\nu} g^{\alpha\beta}  {\Theta^{\gamma\lambda\xi\epsilon\pi}}_{\alpha\gamma\beta}-\mathfrak{D}^{\alpha\beta\mu\nu}R^{\rho\sigma} \bigg\}\nabla_{\lambda}\nabla_{\xi}\delta g_{\epsilon\pi}\nonumber\\
&& +\bigg\{  \mathfrak{C}^{\zeta\mu\nu}g^{\alpha\beta}  {\Theta^{\gamma\lambda\xi\epsilon\pi}}_{\alpha\gamma\beta}\bigg\} \nabla_{\zeta}\nabla_{\lambda}\nabla_{\xi}\delta g_{\epsilon\pi}  \nonumber\\
&&+ \bigg\{ \mathfrak{D}^{\zeta\kappa\mu\nu} g^{\alpha\beta} {\Theta^{\gamma\lambda\xi\epsilon\pi}}_{\alpha\gamma\beta}\bigg\} \nabla_{\zeta}\nabla_{\kappa} \nabla_{\lambda}\nabla_{\xi}\delta g_{\epsilon\pi},
\eea 
which  is clearly of the form (\ref{eq:Sec:deltau_append_modgrav_proto}).
\bibliographystyle{JHEP}
\providecommand{\href}[2]{#2}\begingroup\raggedright\endgroup

\end{document}